\documentclass[3p]{elsarticle}
\newif\ifcomment
\usepackage{parskip}
\usepackage{multicol}
\usepackage{hyperref}
\usepackage{graphicx}
\usepackage{booktabs}
\usepackage{multirow}
\usepackage{subfig}
\usepackage{longtable}
\usepackage{tcolorbox}
\usepackage{lineno,hyperref}
\usepackage{sidecap}
\modulolinenumbers[5]

\journal{Expert Systems with Applications}
\begin{document}

\begin{frontmatter}

\title{Detecting criminal organizations in mobile phone networks}

\author{Emilio Ferrara}
\address{Center for Complex Networks and Systems Research\\
School of Informatics and Computing\\
Indiana University Bloomington\\
919 E. 10th St., Bloomington, IN 47408, USA\\
\texttt{ferrarae@indiana.edu}}

\author{Pasquale De Meo}
\address{Department of Ancient and Modern Civilizations\\
University of Messina\\
viale Annunziata - Polo Universitario, I-98168 Messina, Italy\\
\texttt{pdemeo@unime.it}}

\author{Salvatore Catanese and Giacomo Fiumara}
\address{Department of Mathematics and Computer Science\\
University of Messina\\
viale F. Stagno D'Alcontres 31, I-98166 Messina, Italy\\
\texttt{\{cataneses, gfiumara\}@unime.it}}

\begin{abstract}
The study of criminal networks using traces from heterogeneous communication media is acquiring increasing importance in nowadays society.
The usage of communication media such as phone calls and online social networks leaves digital traces in the form of metadata that can be used for this type of analysis.
The goal of this work is twofold: first we provide a theoretical framework for the problem of detecting and characterizing criminal organizations in networks reconstructed from phone call records. 
Then, we introduce an expert system to support law enforcement agencies in the task of unveiling the underlying structure of criminal networks hidden in communication data.
This platform allows for statistical network analysis, community detection and visual exploration of mobile phone network data. It allows forensic investigators to deeply understand hierarchies within criminal organizations, discovering members who play central role and provide connection among sub-groups.
Our work concludes illustrating the adoption of our computational framework for a real-word criminal investigation.

\end{abstract}


\end{frontmatter}

\linenumbers

\section{Introduction} \label{sec:intro}
We live in a society where ubiquitous connectivity allows millions of users to communicate and enjoy the services provided by the Internet and other communication technologies, now even in mobility, by the technical and commercial success of handheld devices (smartphones, tablets, etc.).
Such type of human communication activities produces a deluge of metadata and digital traces that have been studied to understand inter-connectivity and mobility patterns at scale \cite{onnela2007analysis, onnela2007structure, candia2008uncovering, eagle2008mobile, eagle2009inferring, becker2013human}.
Online social network services such as Facebook and Twitter further increase the amount of information available to describe users' interests, activities and behaviors \cite{Ahn2007, Benevenuto2009, catanese2011crawling, catanese2012extraction, conover2013geospatial, conover2013digital}.
Powerful technologies are although prone to abuse: mobile phone networks and online social media are constantly used to perform or coordinate criminal activities \cite{xu2005criminal, morselli2010assessing}. 
Mobile phone networks can be used to connect individuals involved in criminal activities in real time, often during real-world criminal events, from simple robberies to terror attacks.
Online social media, instead, can be exploited to carry out illicit activities such as frauds, identity thefts or to access classified information. 

Criminal network analysis is pivotal when applied to the investigation of organized crime like terrorism, narcotics trafficking, fraud, etc. \cite{xu2005criminal}.
Criminal organizations are established based on the collaboration of criminals who usually form groups with different roles.
The analysis of a criminal network is thus aimed at uncover the structural schemes of the organization, its operations and, even more importantly, the flow of communications among its members.
In this respect, law enforcement agencies and intelligence agencies often deal with large amounts of raw data gathered from various sources, including phone records and online communication, in order to unveil the network of relations among suspects.
In modern investigative techniques the analysis of phone records represents a first approach that precedes a more refined scrutiny covering financial transactions and interpersonal relations.
For these reasons a structured approach is needed.

The goal of this work is twofold. First, we provide a computational framework based on theoretical foundations and principles from network science, forensic science and statistical analysis to detect and characterize criminal organizations in networks reconstructed from phone communication records. Then, we propose an expert system, called \emph{LogAnalysis}, that implements such framework.

The problem of detecting communities in criminal networks is here formalized as a two-step process: the first step aims at unveiling such communities hidden in larger networks of organic communication involving potentially many individuals, over different time scales; once such criminal organizations are clearly identified, the second step involves the study of the relations existing among the members of the criminal gangs, their communication dynamics, the reconstruction of the their hierarchical relations, to infer the structure of the entire organization and the roles they play therein.

Our expert system implements this computational framework encoding the entire work flow discussed above. \texttt{LogAnalysis} for example automatizes the import of raw phone call records data, the removal of ambiguities and redundancies in data, and the parsing and conversion to a graph format readily available for analysis and exploration.
The data model is designed to improve the quality of the analysis of social relationships observed inside phone call network data through the integration of visualization and social network analysis-based statistical metrics. \texttt{LogAnalysis} implements different state-of-the-art view layouts for promoting fast and dynamic network exploration. It introduces the possibility of analyzing the temporal evolution of the connections among individuals of the network, for example focusing on particular time windows in order to obtain further insights about the dynamics of communications before/during/after particular criminal events. Finally, it provides an unprecedented supervised community detection set of techniques that allows detectives to interact with the community detection process, incorporating expert knowledge to supervise the results and refine the unveiled community structure at different levels of granularity and resolution.

A number of existing tools support network analysis but only some of them have been developed for criminal network investigation.
Related to our work we cite commercial tools like COPLINK \cite{chen2003coplink,xu2005crimenet}, Analyst's Notebook\footnote{i2 - Analysts Notebook. http://www-03.ibm.com/software/products/en/analysts-notebook/}, Xanalysis Link Explorer\footnote{Xanalysis (2014) - http://www.xanalys.com/products/link-explorer/} and Palantir Government\footnote{Palantir government (2014) - http://www.palantir.com/solutions/}. 
Other related prototypes described in academic papers are Sandbox \cite{Brian2006} and POLESTAR \cite{Everett2006}. \textit{LogAnalysis} represents the next-generation criminal investigation expert system in that it introduces significant improvements over these tools, and it provides specific support to detect criminal organizations in network data reconstructed from phone records.

\section{Related Literature} \label{sec:related}

In this section we provide a background on social network analysis, and we survey existing literature in criminal network analysis, with a particular focus on work about communities and communication dynamics. 

Various research streams focus on finding structural properties of criminal networks, including in phone call communication networks \cite{McAndrew99}. Understanding network properties such as the communities present in the network, or the roles that network members play, can help network analysts and police detectives to unveil vulnerabilities and identify potential opportunities to take destabilizing actions to fight criminal organizations. In the following, we discuss relevant work aimed at detecting network communities, discovering their patterns of interaction, identifying central individuals, and uncovering network organization and structure.

\subsection{Background in Social Networks and Models}
Literature about social network models is rooted in social sciences: in the sixties, Milgram and Travers \cite{milgram1967small,Travers1969} analyzed characteristics of real-life social networks, conducting social experiments in the real world, and in conclusion, proposing the well known \emph{small world} model.
They put into evidence that, despite their large dimension, social networks usually show a common property: there exists a relatively short path which connects any pair of nodes within the network.

Another important concept, introduced by Zachary \cite{Zachary1980}, is the \emph{community structure}. 
He analyzed a small real-life social community (i.e., the components of a karate club) and proposed a model which describes the fission of a social network via cuts and divisions in sub-groups.
Nodes in such groups are densely interconnected among each other and weakly interconnected with those belonging to other groups. 

Barab{\'a}si et al. \cite{albert1999diameter, albert2002statistical} introduced a model of network growth which can be applied to friendship networks, the World Wide Web, communication networks, etc. 
The authors proved that such networks share the same dynamics of growth, called \emph{preferential attachment}: new nodes tend to preferentially connect to existing nodes with high degrees rather than lower degree ones.
This characteristic yields to the emergence of \emph{scale-free distributions} in the degree of the nodes, allowing for the presence of hubs and spokes in the network. 

The concurrent presence of the small world effect, the emergence of a community structure and the preferential attachment mechanism are the three crucial ingredients that characterize the structure of social networks \cite{ferrara2011topological}.

\subsection{Criminal Network Analysis and Community Structure} 
\label{sub:community-structure}
In the latest years, the academic community working on the application of social network analysis (SNA) to intelligence and study of criminal organizations has been constantly growing.
One of the main contributions in this field is due to Sparrow \cite{sparrow91}, who focused on the application of SNA in order to identify the vulnerabilities of different types of criminal organizations.
He highlighted three key aspects of Criminal Network Analysis (CNA), namely: i) the importance of SNA in order to analyze information; ii) the potential of intelligence when applied to the analysis of the networks; and, iii) the common results obtained from the collaboration of the two sectors.
Sparrow also introduced the following definitions: i) dimension --- the Criminal Networks (CNs) may have up to thousands elements; ii) incompleteness ---  criminal or terroristic networks are inevitably incomplete due to the fragmentary or erroneous information available; iii) undefined borders --- it is difficult to determine all the relations of each member; and, iv) dynamism --- new connections necessarily imply an evolution of the structure of the network.

Starting from Sparrow's work, several authors tried to augment the superposition between the two fields by analyzing the CNs with the instruments typical of SNA: this is the case, for example, of the analysis Baker and Faulkner \cite{BakerFaulkner93} carried out on illegal networks in the field of electric plants, of Klerks' study \cite{Klerks01thenetwork} of criminal organizations in Netherlands, and the network analysis of Iranian government carried out by Deckro and Renfro \cite{Renfro01}. 
In 2001, Silke \cite{Silke01} and Brannan et al. \cite{Brannan01} examined the state of research in the field of terrorism and documented some cases in which it was lacking and empiric.
Arquilla and Ronfeldt \cite{Arquilla01} summarized the preceding work and introduced the concept of \emph{NetWar} and its applicability to terrorism.
In particular, they drew attention to the differences existing between the analysis of social and criminal networks, and highlighted the usefulness of research in these fields in order to understand the nature of criminal organizations.
Notwithstanding the fact that the framework proposed by Arquilla and Ronfeldt provided a novel method to conceive network analysis, these authors received disapprovals because their approach was considered purely theoretical.
Before 2001-09-11 one of major criticisms came from Carley, Reminga e Kamneva \cite{Carley2006}, concerning the initiatives for destabilizing the dynamic terroristic networks.

In 2006, Krebs \cite{Krebs02mappingnetworks} applied network theories to the analysis of the Al Qaeda cell responsible of the 2001-09-11 attack.
That work started a series of academic papers in which SNA has been directly applied to real cases, differently from previous research which was applied to artificial data or networks.
Krebs' paper is still one of the most cited works in the field of the application of SNA to criminal networks and inspired a number of SNA applications used by intelligence agencies for the counter-terrorism war.

Several studies have been conducted in order to investigate the community structure of real and online social networks \cite{fortunato2010community}.
The problem of finding communities in a network is often formalized as a clustering problem.
There is one widely adopted approach to solve this problem, based on the concept of \emph{network modularity}, which can be explained as follows: let consider a network, represented by means of a graph $G=(V,E)$, which has been partitioned into $m$ communities; its corresponding value of network modularity is 
\begin{equation} \label{eq:qmod}
	Q= \sum_{s = 1}^m \left[\frac{l_s}{\mid E \mid} - \left(\frac{d_s}{2\mid E \mid}\right)^2\right]
\end{equation} 
assuming $l_s$ the number of edges between vertices belonging to the $s$-th community and $d_s$ is the sum of the degrees of the vertices in the $s$-th community.
High values of $Q$ imply high values of $l_s$ for each discovered community, yielding to communities internally densely connected and weakly coupled among each other. 

The network modularity is therefore used as fitness function to solve an optimization problem: several methods exist, including the Girvan-Newman algorithm and its optimized variants \cite{girvan2002community, newman2004finding, newman2006modularity, blondel2008fast}.
The goal of such strategies is that of producing a network clustering that exhibits a high network modularity. 
Although such methods are usually efficient, two limitations exist: first, the modularity function carries a resolution limit \cite{fortunato2007resolution} preventing the detection of communities smaller than an intrinsic scale determined by the network size and its inter-connectivity; moreover, such techniques produce hard partitioning of the networks thus assigning each node to one and only one community. 

Strategies to work around both limitations exist, and recently some approaches have been proposed to discover overlapping communities \cite{palla2005uncovering, xie2011overlapping}, in order to allow nodes to belong to different communities.
Our framework is based on network modularity maximization methods and therefore the potential impact of such limitations will be discussed further into details later on. 
We will highlight that the ability that \textit{LogAnalysis} provides to supervise the community detection process yields to more refined communities with respect to automatic methods that may suffer from the above-mentioned limits.

The strength of \textit{LogAnalysis} consists in the adoption of several statistical and interactive visualization layout techniques that improve network analysis while highlighting different aspects and features of the considered network and identifying and visualizing community structures.

\section{\textit{LogAnalysis}: main features} \label{sec:main-features}

In this section we summarize the main features of \emph{LogAnalysis} including metrics  and visualization layouts. 

\subsection{Network metrics} \label{sub:NetworkMetrics}
Members of criminal networks dynamically modify their relations with other members of the network thus resulting in a change of their role and importance.
A series of centrality measures typical of the Social Network Analysis can help in capturing these changes.

These statistics are used to filter the network view based on specific node value and highlight their position inside the network.

\textit{Degree centrality} is defined as the number of direct links a node has.
A node with a high degree can be seen as a hub, an active node and an important communication channel.

\textit{Betweenness centrality} measures the extent to which a particular node lies between other nodes in a network. 
These intermediate elements may wield strategic control and influence on many others. The core issue of this centrality measure is that an actor is central if he lies along the shortest paths connecting other pairs of nodes.
An individual with a high betweenness may be a gatekeeper in the network.
A gatekeeper criminal should often be targeted for removal because the removal may destabilize a criminal network or even cause it to fall apart \cite{Carley2006}.

\textit{Closeness centrality} is the inverse of the sum  of the shortest paths (geodesics) connecting a particular node to all other nodes in a network. 
The idea is that an actor is central if he can quickly interact with all the others, not only with his first neighbors \cite{Newman_2005}.
In the context of criminal networks, this measure highlights entities with the minimum distance from the others, allowing them to pass on and receive communications more quickly than anyone else in the organization.
For this reason, the adoption of the closeness centrality is crucial in order to put into evidence inside the network, those individuals that are closer to others (in terms of phone communications).
In addition, high values of closeness centrality in this type of communication networks are usually regarded as an indicator of the ability of the given actor to quickly spread information to all other actors of the network.

\textit{Eigenvector centrality} is another way to assign the centrality to an actor of the network based of the idea that if a node has many central neighbors, it should be central as well. 
This measure establishes that the importance of a node is determined by the importance of its neighbors.
In the context of telecom networks, eigenvector centrality is usually regarded as the measure of influence of a given node. High values of eigenvector centrality are achieved by actors who are connected with high-scoring neighbors, which in turn, inherited such an influence from their high-scoring neighbors and so on.
This measure well reflects an intuitive important feature of communication networks that is the influence diffusion.

\textit{Clustering coefficient (transitivity)} of a graph measures the degree of connectedness of a network.
High clustering coefficients mean the presence of a high number of triangles in the network.
Is well-known in the literature \cite{wasserman1994social} that communication networks show high values of clustering coefficient since they reflect the underlying social structure of contacts among friends/acquaintances.
Moreover, high values of local clustering coefficient are considered a reliable indicator of of nodes whose neighbors are very well connected and among which a substantial amount of information may flow.

\subsection{Network layouts}
\textit{LogAnalysis} has been developed as a tool to help forensic detectives in the analysis of phone log records by means of a network representation. We adopted different state-of-the-art view layouts for promoting fast exploration and discovery of the analyzed networks.

It allows to analyze the relational structure of a criminal networks and to unveil the mechanisms of communications among its members.
Various types of relations can be categorized by identifying those members who occupy central positions, those who play a key position in the communication flows among various groups (clans), etc.
The application also enables to study the temporal evolution of the criminal network and to highlight some crucial information regarding the dynamics of the links in concurrence with criminal events.

\subsubsection*{Node-link}
\begin{figure}[!h]
\centering
\includegraphics[width=1.00\columnwidth]{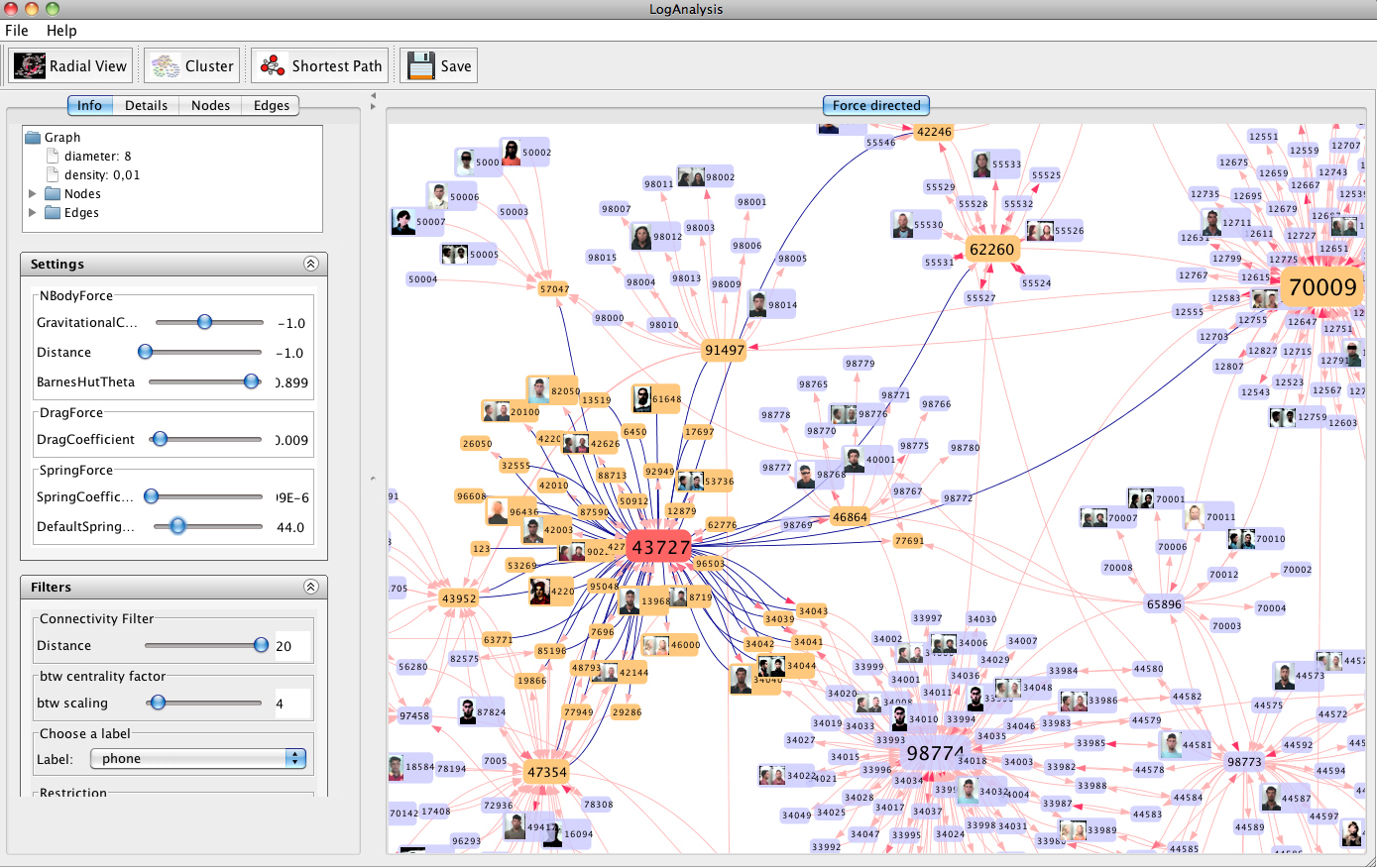} 
\caption{\emph{Log Analysis} interface and force-directed layout. This figure shows the criminal network resulting from a case study of 543 nodes and 1229 edges. The node labeled in rad has been selected by the user. The nodes labeled in yellow are those at distance 1 from the selected node.}
\label{fig:logAnalysisUI}
\end{figure}

Phone calls logs infer a social network. The tool mainly employs the node-link representation in order to visualize networks in which node was created for each unique cell phone, and an edge was created for each phone call.
This results in a social network as shown in Figure \ref{fig:logAnalysisUI}. 

To increase the readability of the network, when the mouse is passed over a node, first order connections of the node are highlighted.
Moreover, it is possible to set the distance-based filter in order to represent only the nodes which fall within a given distance from the selected node.
\textit{LogAnalysis} also includes panning and zooming, and it implements the search by means of textual keys with the subsequent highlighting of nodes matching the query criteria.

\subsubsection*{Radial Tree}
\begin{figure*}[!h]%
	\centering
	\includegraphics[width=\columnwidth]{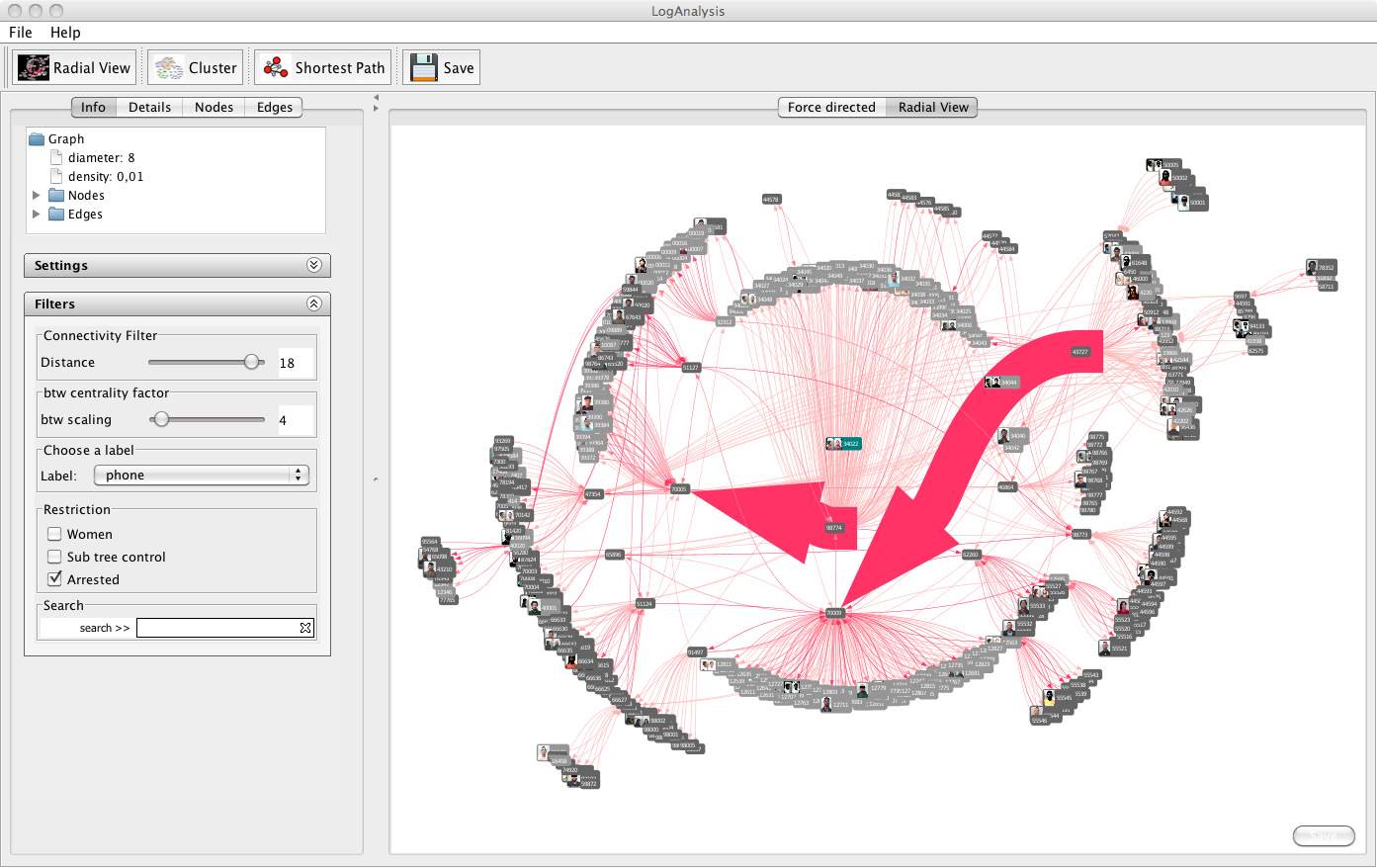}%
	\caption{Example of Radial View layout. The node selected by the analyst is central in this visualization. The thickness of the edges connecting pairs of nodes is proportional to the amount of communication flowing between those pairs.}%
	\label{fig:radial}%
\end{figure*}

As shown in Figure \ref{fig:radial} Radial tree layout allocates the elements of a graph in radial positions and defines several levels upon concentric circles with progressively increasing radii.
The algorithm \cite{yee2001animated} also puts nodes in radial positions but gives the possibility of varying positions while preserving both orientation and order.

According to that technique, a selected element is placed at the center of the canvas and all the other nodes are subsequently placed upon concentric circles with radii increasing outwards.
This visualization strategy is instrumental in the context of the forensic analysis because it allows to focus the attention of detectives on a suspect, and to have a close look to its connections.

Nodes lying on the circumference of concentric circles, centered on that node, could be also progressively displaced from the selected one.
Moreover, edges are visualized by using different thickness, calculated with respect to the number of calls among the given connected nodes.

\subsubsection*{Dinamics analysis}
Phone call networks are not static and the structure of the network could change over time. So it is crucial for investigators ``filtering'' an analyzing the the social dynamics of the network with respect to specific temporal constraints.
Our tool provide three temporal analysis features, shown in Figure \ref{fig:DinamicFeatures}, that heightens these capability:

\begin{figure}[!t]
\centering
\subfloat[][\emph{Time filter}.]
{\includegraphics[width=0.98\columnwidth]{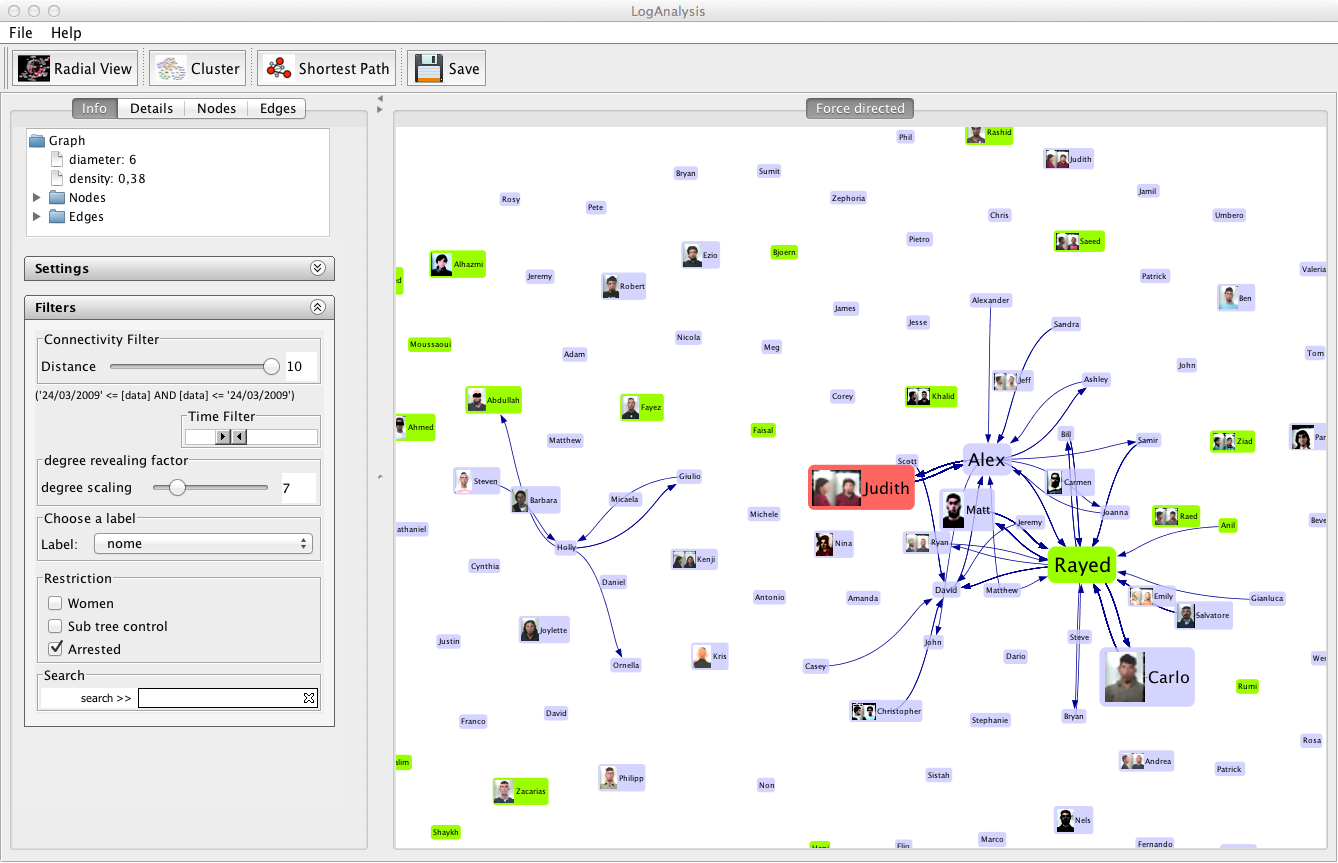}} \\
\subfloat[][\emph{Time flow analyzer}.]
{\includegraphics[width=0.50\columnwidth]{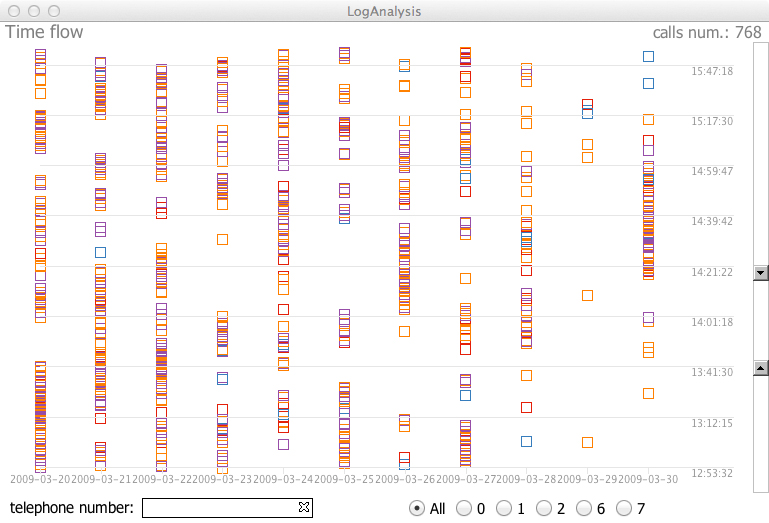}}
\subfloat[][\emph{Stacked histogram}.]
{\includegraphics[width=0.50\columnwidth]{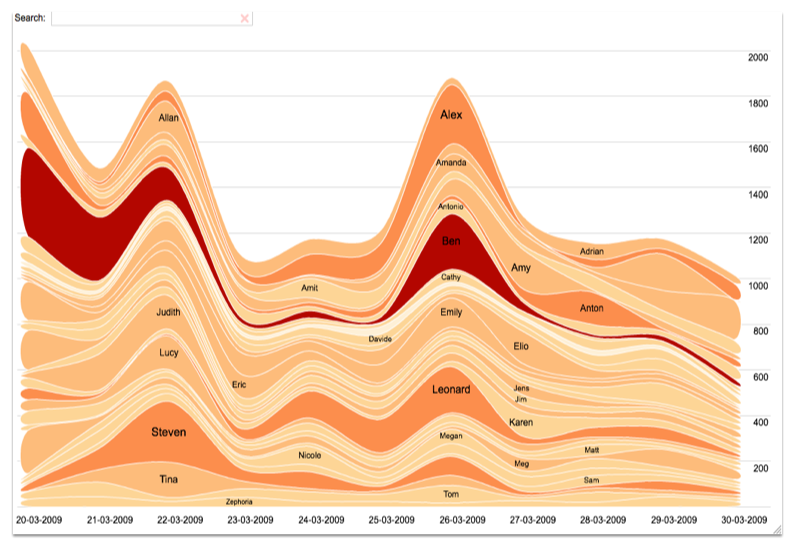}}
\caption{(a) The Time Filter feature allows to investigate the network structure evolution. Nodes are dynamically engaged or detached according to the time range slider. (b) The Time Flow scatterplot is helpful to consider the time-dependence of events (i.e., phone calls) in a specific time window and it is crucial to highlight phone call cascades during criminal events. (c) The Stacked Histogram is helpful to visually summarize the communications among actors elapsed in a temporal interval.}
\label{fig:DinamicFeatures}
\end{figure}

i) \textit{Time filter}. It is possible to select a time slice by using a slider. The structure of the network is filtered accordingly, removing all the edges representing connections (i.e., phone calls) which did not take place in that specific time window, and insulating (or hiding) those nodes not involved in the network at that given time. Modiying the time interval, nodes involved are automatically ``engaged'' or detached and are attracted or rejected inside/outside the network.

ii) \emph{Time Flow analyzer} considers each single phone call as an \emph{event} in a scatterplot. The days are on the x-axis and the hours on the y-axis. The colors of nodes are determined by the type of communication (i.e., sent/received calls and SMS and other type of communications, etc.). 
User can zoom in/out the time interval using a range slider to obtain additional insights about connections of events and
query the data about specific key world.

iii) \textit{Stacked histogram}. In this visualization each node in the network is assigned a stack. 
The thickness of each stack is according to the nodes degree at the time on the horizontal axis.
This feature is helpful to get a picture of the phone call activity of the set of suspected elapsed during a specific time window ed in particular before, during and after criminal commission event.
It is helpful to some interesting discoveries. For instance, why after the peak cell phones not contact each other? Any why, did the activity increase in a specific date?

\section{Criminal Network Community Detection} \label{sec:community}
A criminal network is a special kind of social network with emphasis on both secrecy and efficiency. Such networks are intentionally structured to ensure efficient communication among members without being detected \cite{Gniadek2010}.
Knowledge about the criminal network structure is crucial to the investigators in order to reveal the functional or operational nature of the organization.

Typical criminal network information structures that emerge during investigations include hierarchical structure \cite{sageman2004}, cellular structure \cite{Nomani2012} comprised of cohesive subgroups connected by bridges, and flat structure \cite{Krebs02mappingnetworks}. These structures are emergent and evolving as the criminal network is modeled incrementally.

One of the most relevant features of graphs representing real systems like criminal networks is the community structure, or clustering. 

The main goal of community detection in criminal networks (in particular, in phone call networks) is the identification of groups (or, clans) and their structures thanks to information coded in the topology of the corresponded graph.

In this section we will discuss how we approached the problem algorithmically and in terms visualization layout. To detect the community structure of the criminal network discusses as case study in this paper we use our tool \textit{LogAnalysis}.
This framework includes two strategies to detect and explore communities: i) Girvan and Newman algorithm \cite{girvan2002community} (in the following, GN) and ii) a variant based on modularity optimization, known as Newman's algorithm \cite{newman2004fast} fast enough to support interactive real-time adjustments.

The simple idea behind the GN algorithm is to identify those edges that interconnect nodes belonging to different clusters and progressively remove them, so that the clusters are disconnected and the community structure emerges. The identification of bridge edges can be obtained by various means. In the case of GN, the algorithm adopts the edge betweenness centrality.

The following steps describe in detail how the GN algorithm works: \emph{(i)} the edge betweenness of all the edges is computed; \emph{(ii)} the edge with the highest value of edge betweenness is removed; \emph{(iii)} the edge betweenness is computed for the new configuration and, \emph{(iv)} the algorithm is repeated going back to step \emph{(ii)}.

The edge betweenness centrality is computed in a $O(mn)$ time, $m$ being the number of the edges and $n$ the number of nodes.
It has to be repeated $m$ times, so the worst computational cost is $O(m^2n)$, or $O(n^3)$ for sparse graphs.
Note that, although for large networks such high computational cost makes this solution often unfeasible, for criminal networks constituted (most often) by hundreds or at most thousands of nodes, this algorithm works well.

\textit{LogAnalysis} visually presents the communities identified by GN via a force-directed node-link layout  \cite{fruchterman1991graph}.
The deletion of an edge affects the structure of the network, iteration after iteration, and the network is represented accordingly: deleted edges are depicted as transparent.
The number of the edges to be deleted can be chosen interactively.
Finally, nodes are colored according to the cluster they belong to.

\begin{figure}[!h]
\centering
\subfloat[][\emph{Phone call network of 148 nodes and 210 edges}.]
{\includegraphics[width=.48\columnwidth]{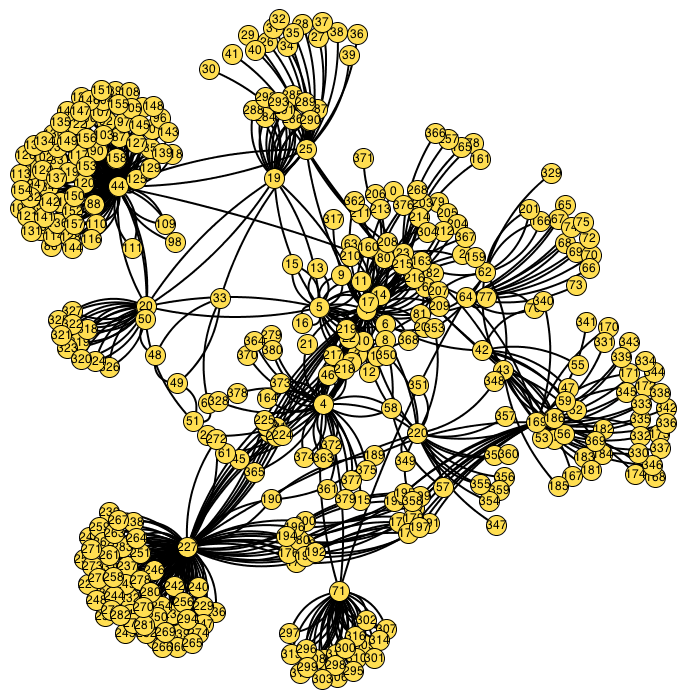}} \quad
\subfloat[][\emph{Clusters detected after 46 edges deleted}.]
{\includegraphics[width=.48\columnwidth]{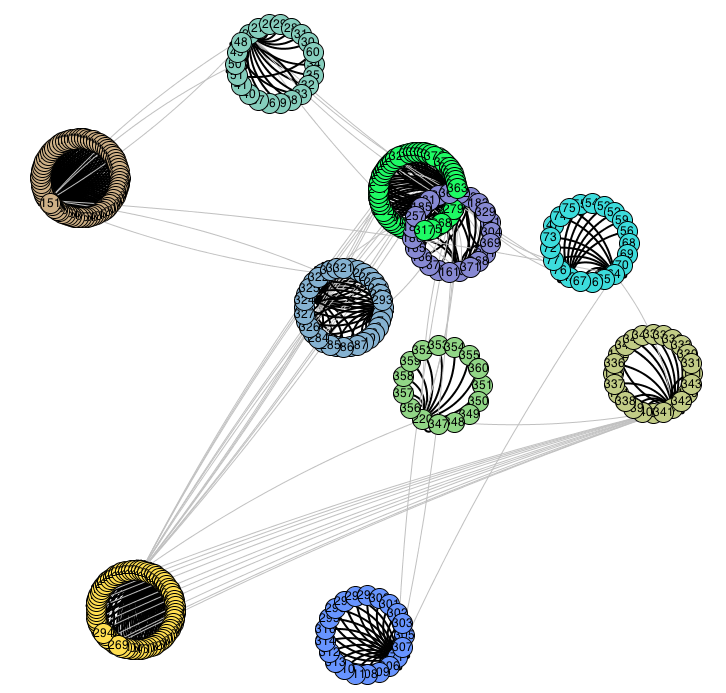}}
\caption{Community detection using the Girvan Newman algorithm and the Fruchterman-Reingold layout. The sequence shown: (a) a phone call networks of 148 nodes and 210 edges (b) clustered view after 46 edges deleted. In this configuration modified force-directed algorithm visually present communities in circular layout.}
\label{fig:EsempioClustering}
\end{figure}

Figure \ref{fig:EsempioClustering}$(a)$ illustrates the typical structure of a network representing the phone calls network of 148 nodes and 210 edges, according to the node-link layout.
Figure \ref{fig:EsempioClustering}$(b)$ shows the network after 46 iterations of the GN algorithm: 10 communities have been detected. This configuration is a modified force-directed layout in which community members are visualized using a circular layout.

This characteristic of the GN algorithm is particularly well suited for the analysis of criminal networks: when the most central edges are progressively deleted, intermediate structures emerge, and an appropriate level of clustering can be determined.

The second method used in \textit{LogAnalysis}, the Newman's fast algorithm, is a variant of GN aiming to maximize the network modularity function as described in Section \S\ref{sub:community-structure} by means of a greedy strategy.
It is a hierarchical clustering method in which groups of nodes are progressively aggregated in order to form larger communities whose modularity increases after the aggregation.
At the first step, $n$ clusters are considered, each composed of a single node.
Edges are added one by one during the procedure.
The modularity of the partitions is computed by taking into account the complete topology of the network.
By adding the first edge to the set of disconnected nodes, the number of groups is decreased to $n-1$ so that a new partition is obtained.
At each step of the algorithm the edges to be added are chosen so that the partition obtained results in an increase, or at least the minimal decrease, of the modularity with respect to the previous configuration.
At each iteration, the variation $\Delta Q$ of modularity is to be computed as a result of the fusion of two any communities belonging to the running partition so to allow to choose the best resulting partition.
The algorithm requires $n-1$ iterations, therefore its computational complexity is $O ( ( m + n )n)$, or $O(n^2)$ in the case of a sparse graph.
As a consequence, the community detection is feasible in the case of networks larger than those which can be tackled using the GN algorithm.

\begin{SCfigure}[][!h]
	\centering
	\includegraphics[width=0.75\textwidth]{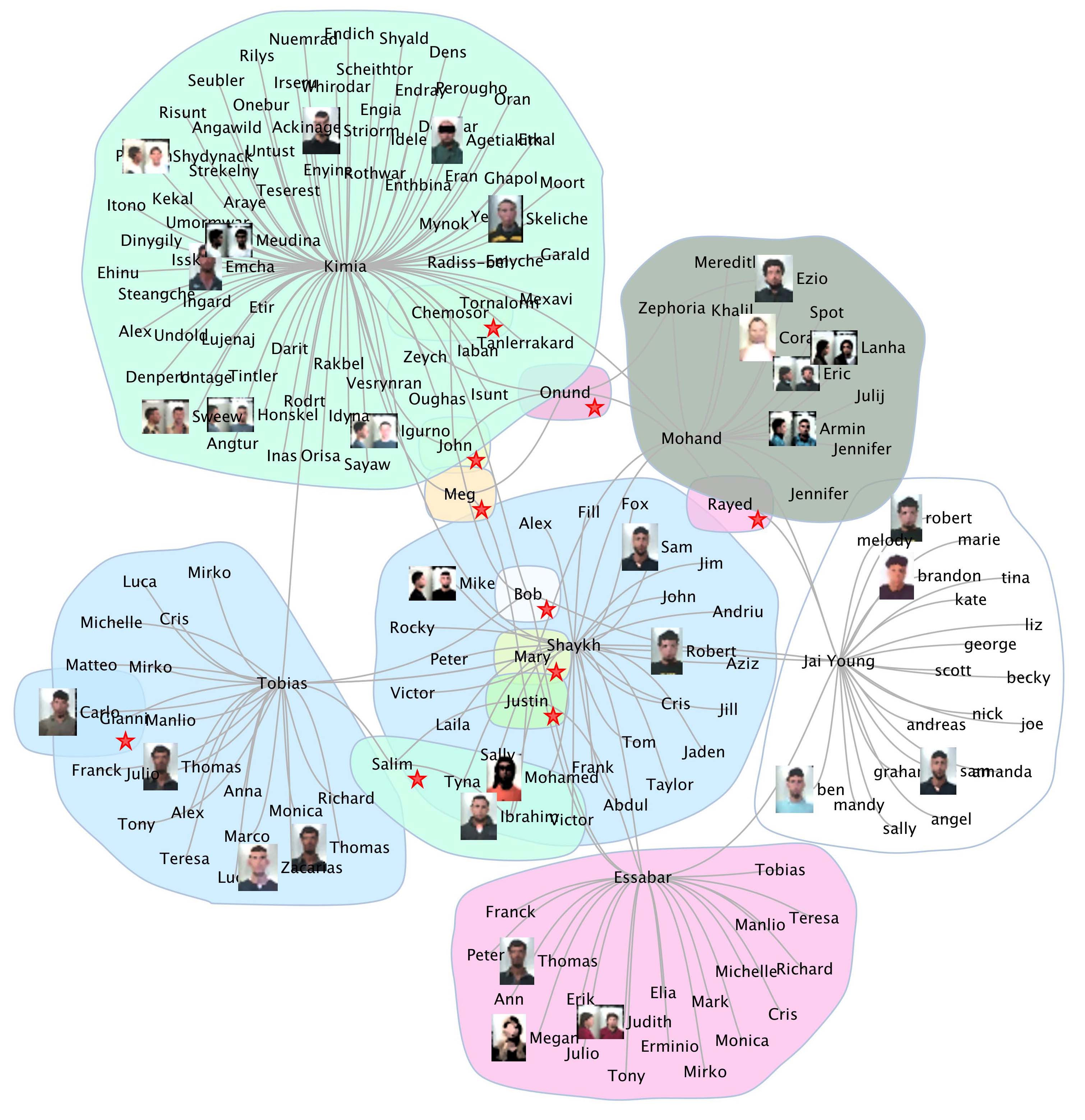}
	\caption{Community layout: Newman's fast algorithm \cite{newman2004fast}. The algorithm finds fourteen communities, eight of which are collapsed into a single node (For privacy reasons, photos have been anonymized).}
	\label{fig:Newman}
\end{SCfigure}

To visually present the Newman's algorithm results in \textit{LogAnalysis}, community are shown within ``convex hulls'' (like in Vizster \cite{heer2005vizster}).
Additional forces separate the communities avoiding their overlapping.
Besides the visualization of communities inside the hulls, it is possible to filter and navigate the network by compressing the clusters around their most representative (i.e, central) nodes.
Figure \ref{fig:Newman} shows a Newman community detection on a 223-node network. 

Generally speaking, community detection methods based on modularity optimization are imperfect: some detected clusters can be larger with respect to the clans really existing in the network. 
This effect can be related to the resolution limit \cite{fortunato2007resolution} mentioned in Section \S\ref{sub:community-structure}.
To overcome this problem, GN and Newman algorithms are combined with a parameter so that users can tune the state of clustering at an given granularity.
Analysts can split/merge the communities into smaller/bigger groups and can choose configurations that make more sense for their analysis. This capability is especially useful with dynamic networks such as telephone call ones, in which interpersonal relationships and the organization structure may changes over time.

\begin{figure}[!h]
\centering
\subfloat[][\emph{Sample network of 18 nodes and 30 edges where 5 communities have been detected}.]
{\includegraphics[width=.48\columnwidth]{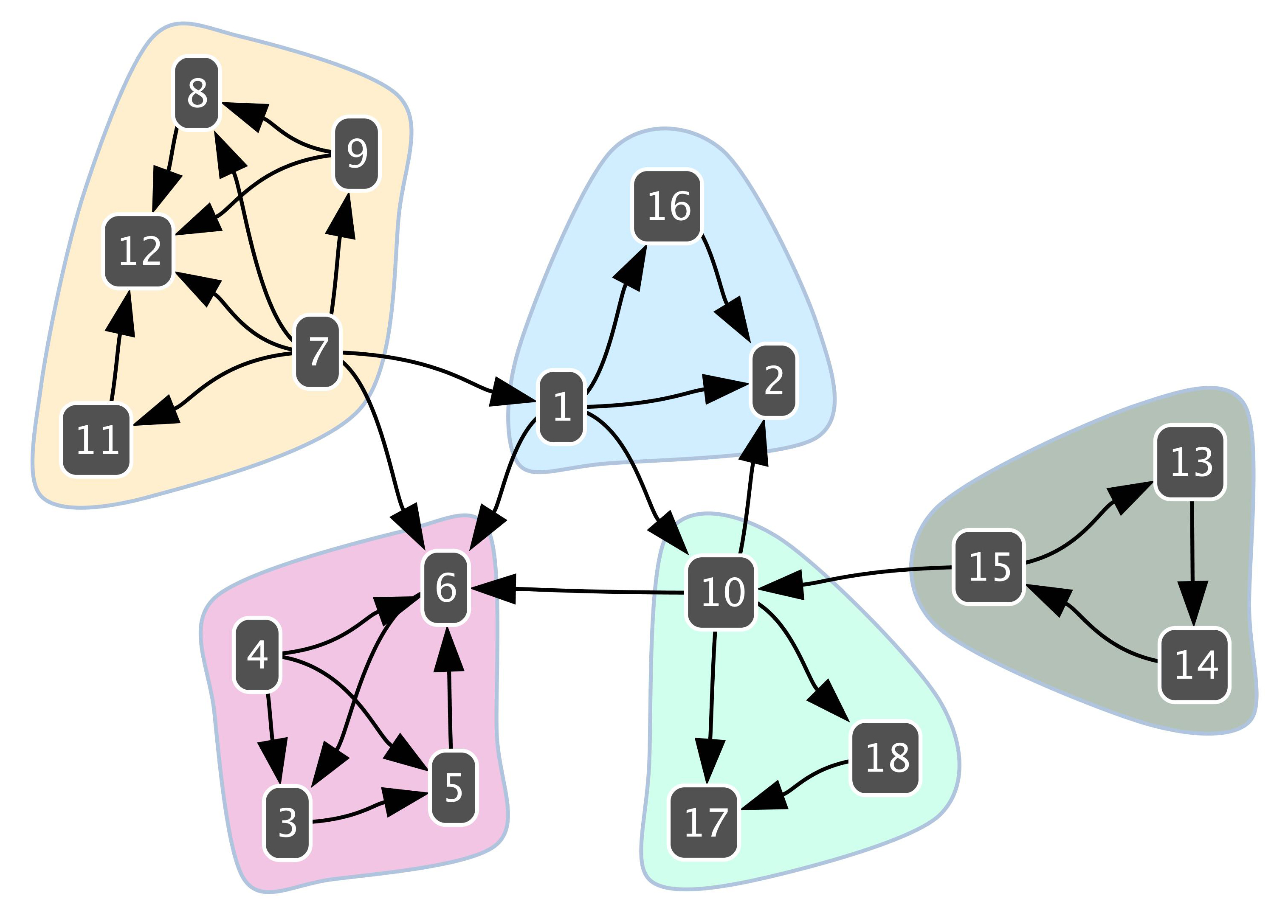}} \quad
\subfloat[][\emph{Selected community with 4 nodes. The node labeled with a red star indicates a connection to one or more communities currently collapsed}.]
{\includegraphics[width=.48\columnwidth]{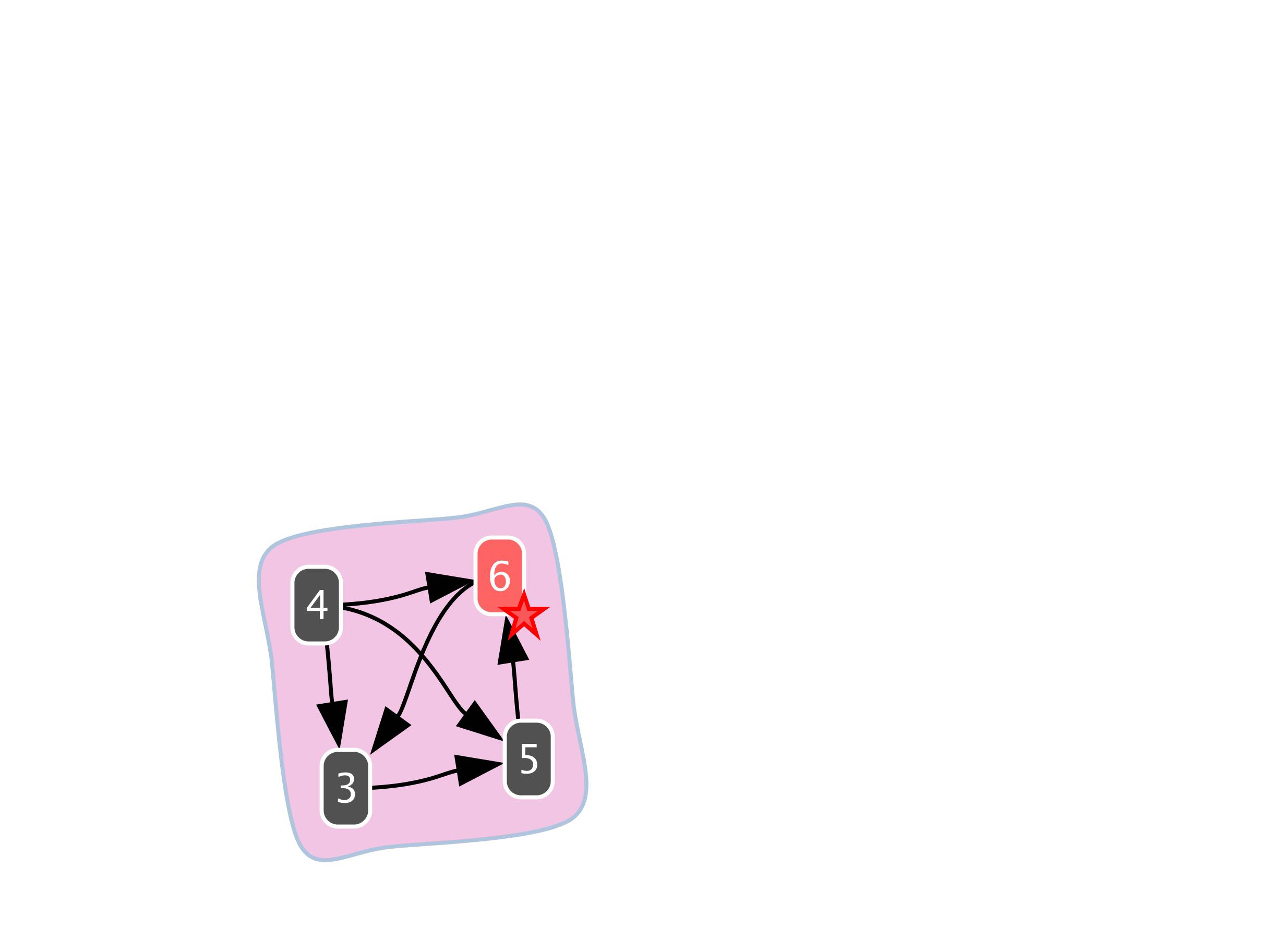}} \\
\subfloat[][\emph{Configuration obtained setting to 1 the inter-community hop parameter. Red stars indicate convex hulls representing collapsed communities}.]
{\includegraphics[width=.48\columnwidth]{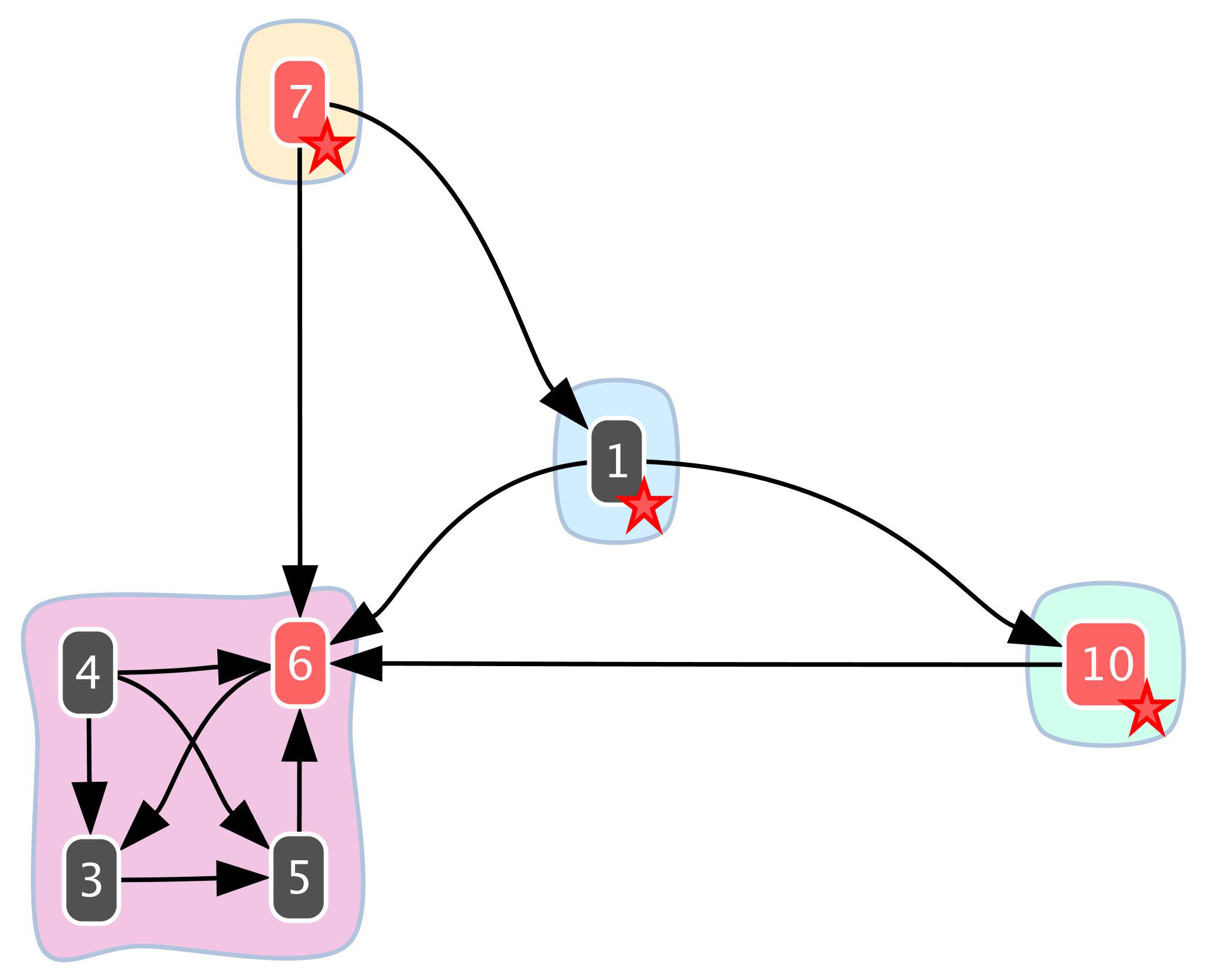}} \quad
\subfloat[][\emph{By setting the inter-community hop to 2, some communities are automatically exploded, while one remains collapsed}.]
{\includegraphics[width=.48\columnwidth]{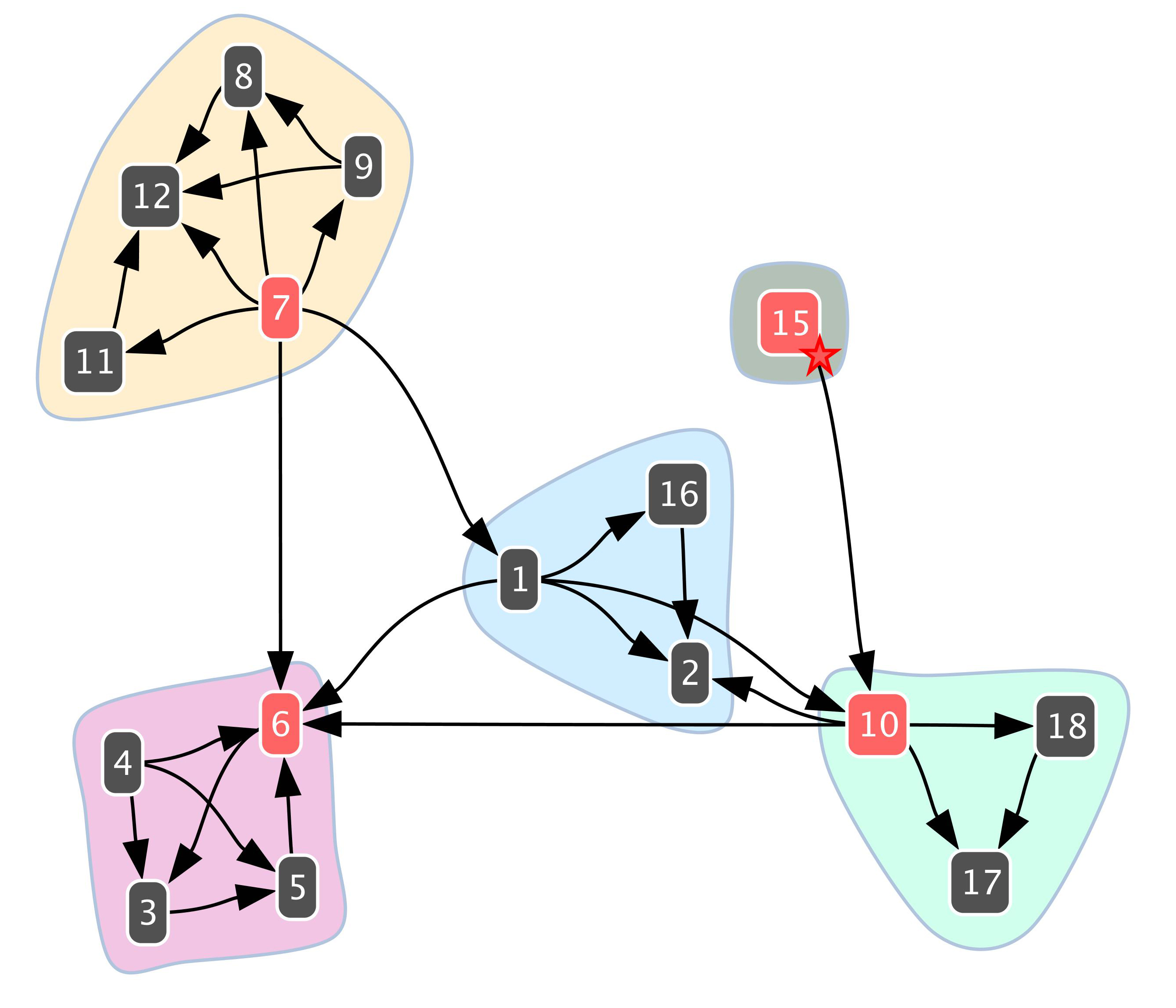}}
\caption{Example of community detection with the Newman algorithm, visualization and interactive exploration.}
\label{fig:ConvexHullSample}
\end{figure}

Another feature of \textit{LogAnalysis} is the possibility of interactively analyze the communities detected by the Newman algorithm. One example of this type of investigation is illustrated in Figure \ref{fig:ConvexHullSample}. In this small network of 18 nodes and 30 edges, 
the algorithm detected 5 communities (see Figure \ref{fig:ConvexHullSample}(a)). 
By setting to zero the parameter that tunes the number of inter-community hops, and selecting the convex hull of a given cluster (for example, the one containing four nodes, like in Figure \ref{fig:CaseStudyNewman}(b)), the graph is filtered and collapsed accordingly: Figure \ref{fig:CaseStudyNewman}(c) shows how the specific cluster connects to the others (note that collapsed communities are identified by a red star and labeled with the id of the node(s) distant one hop from the selected node in the selected cluster). When the number of inter-community hops is set to 2, some of the clusters are automatically exploded (see Figure \ref{fig:CaseStudyNewman}(d)) indicating that such communities are reachable in one hop from the selected one, while some others remain collapsed because their members are farther away from the selected community. 

The set of techniques presented above simplify greatly the analysis of criminal networks inferred from (possibly large) phone call data. 
Our system allows to achieve a trade-off between granularity of the information presented in the visual interface, and the ability for the analyst to explore large networks and the criminal communities therein exposed. In the next section we focus on the characteristics of criminal network by means of a case study reconstructed from a real criminal investigation supported by our expert system. 

\section{A case study} \label{sec:acasestudy}

Real police investigations have been successfully carried out supported by \textit{LogAnalysis}. 
In this Section we report a case study whose results were obtained during a forensic investigation.
Although for sake of privacy protection some information is obfuscated, in the following we will drive the reader through all steps of the criminal investigation carried out by means of our framework. 

\subsection{The initial configuration}
In this case, some people allegedly belonged to a criminal network.
Among the available data about the structure of the criminal organization, phone logs undoubtedly convey the most important information that detectives use in order to verify the existence of interpersonal relationships and the communication flow.

\begin{figure}[!h]
\centering
\subfloat[Case study network in node-link layout.]{
\includegraphics[width=0.80\columnwidth]{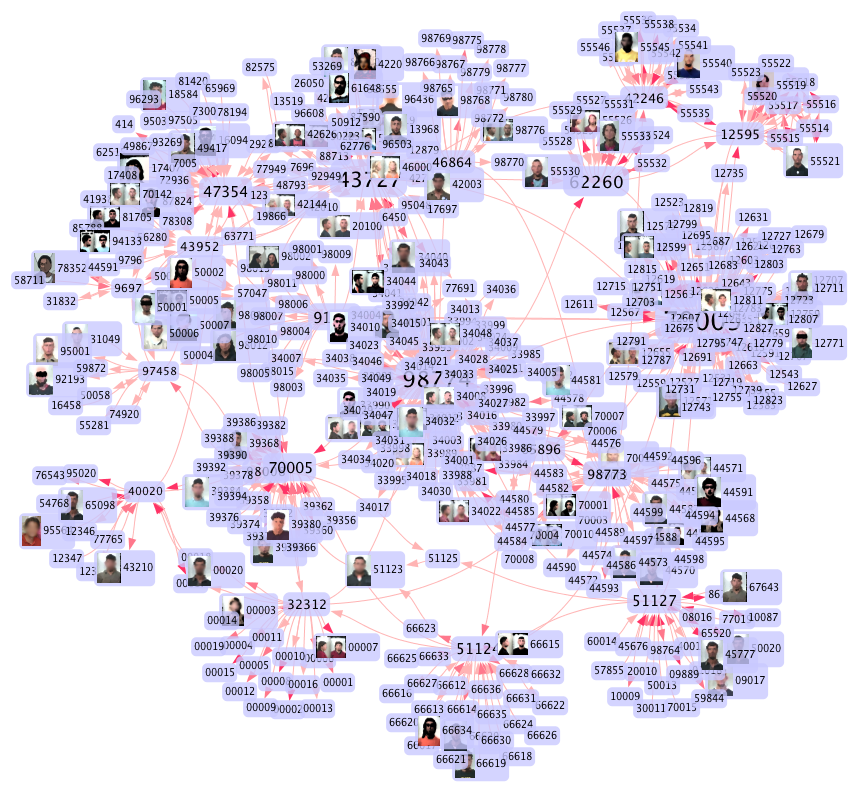}}\\
\footnotesize {
\subfloat[Overall Metrics.]{%
\begin{tabular}{lc}
\toprule
Network Metric & Value \\
\midrule
Network type & directed \\
Vertices & 381 \\
Edges & 428 \\
Connected components & 1 \\
Self-Loop & 0 \\
\midrule
Maximum Geodesic Distance & 7 \\
Average Geodesic Distance & 3.898 \\
Network Density & 0.006 \\
\midrule
RecordSet & 15,845 \\
SMS & 6342 \\
MMS & 133 \\
Voice calls & 7,334 \\
Internet & 1,180 \\
Others & 856 \\
\bottomrule
\end{tabular}}%
}
\quad
\footnotesize{
\subfloat[Centrality measures of top 15 vertices in.]{%
\begin{tabular}{rccc}
\toprule
vertex & degree & btw centr. & page rank \\
\midrule
134 & 845 &  5741.467	& 6.941 \\
32 & 710 & 5870.000	& 5,718 \\
18	& 532	& 13130.767	& 6.150 \\
31	& 358	& 14245.000	& 14.183 \\
91	& 349	& 12647.173	& 12.756 \\
94	& 220	& 31622.172	& 19.896 \\
106	& 211	& 19505.405	& 8.407 \\
37	& 188	& 16559.613	& 7.458 \\
102	& 163	& 28694.821 & 35.803 \\
128	& 157	& 8293.357	& 9.812 \\
124	& 152	& 5515.127	& 5.397 \\
289	& 133	& 21104.667 & 29.204 \\
25	& 130	& 4637.467	& 5.598 \\
16	& 110	& 3224.286	& 4.498 \\
69	& 105	& 2742.500	& 4.117 \\
\bottomrule
\end{tabular}}
}
 \caption{$(a)$ Network visualization in node-link layout of the entire cell phone log data set composto da 381 nodes and 428 in 15 days of activities. Each node is a unique cell phone, and each edge is a relationship (calls, SMS, MMS, etc.) between them. $(b)$ Overall metrics and $(c)$ centrality measures of the top 15 vertices of the case study network.}
 \label{fig:Metrics}
\end{figure}

The initial configuration of the network representing the phone call connections is shown in Figure \ref{fig:Metrics}.
The network has been obtained from the processing of the log files containing the phone call traffic during a period of fifteen days among some people allegedly belonging to a criminal association responsible of a series of robberies, extortions and drug illicit trafficking.
\textit{LogAnalysis} may also automatically expand metadata on actors of the network, whether available: in this case (obfuscated) mugshots, and other metadata (e.g., criminal records, etc.), are autonomously extracted by consulting other internal police databases.
In addition, for anonymization purpose, phone numbers are here replaced by numerical IDs. 
Information concerning the relational structure and some important statistical metrics are shown in Figure \ref{fig:Metrics}.

From the analysis of phone contacts among some people in a given time interval it is also possible to unveil the most important links in terms of frequencies of relations and flow of information.
Links do not refer to the same type of relations and therefore it is important to improve the analysis starting from the community detection.
Crucial is the ability to gain as much information as possible from the topology of the network and then ascertain the details.

\subsection{Finding subgroups}

\begin{figure}[!h]
 \begin{minipage}[t]{.68\linewidth}\vspace{0pt}
 	\subfloat[][\emph{Case study network after the GN algorithm. \\ 16 communities have been detected}.]
	{\includegraphics[width=.99\columnwidth]{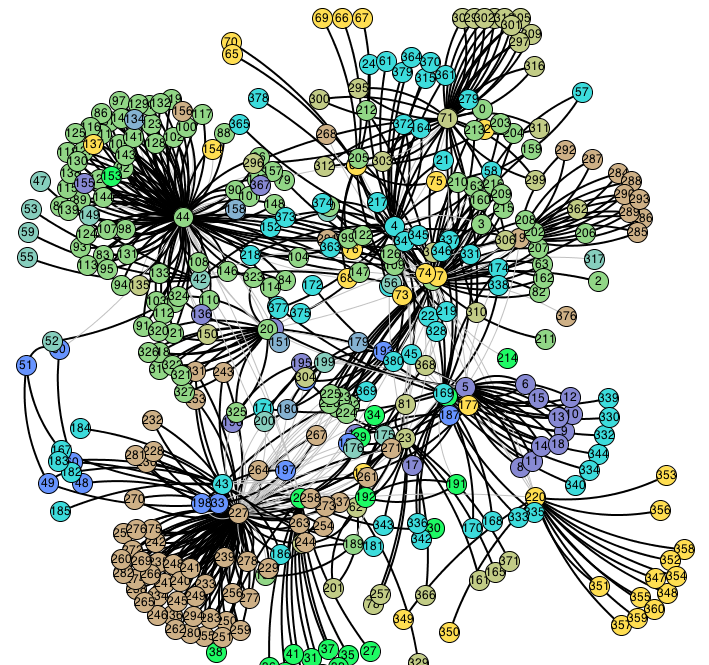}}
 \end{minipage}
 \hfill
 \begin{minipage}[t]{.32\linewidth}\vspace{0pt}\raggedright
  \begin{minipage}[t]{\linewidth}\vspace{0pt}\raggedright
   	\subfloat[][\emph{Clustered view. Nodes of the same community form macro-nodes visualized with a circular layout}.]
	{\includegraphics[width=1.0\columnwidth]{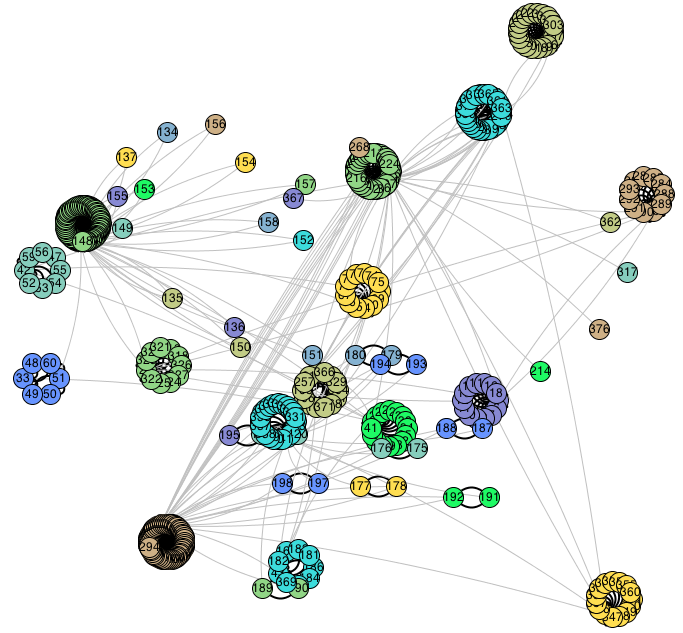}} \\
	\subfloat[][\emph{Macro-nodes zoom reveals intra-community relationships}.]
	{\includegraphics[width=1.0\columnwidth]{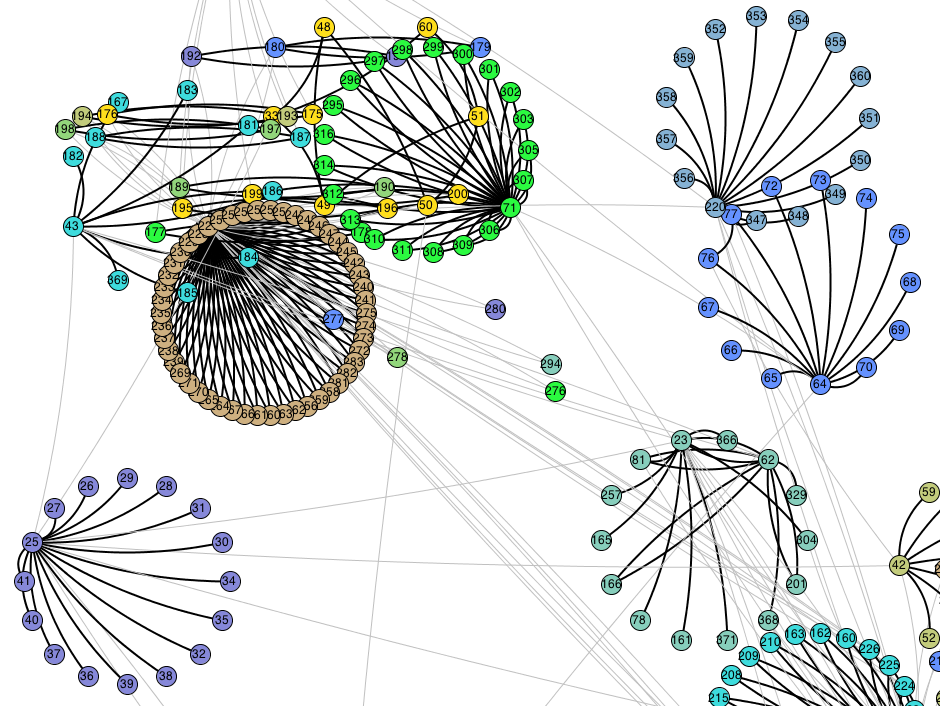}}
  \end{minipage}
  \begin{minipage}[t]{\linewidth}\vspace{20pt}\raggedright
	\end{minipage}
\end{minipage}
 \caption{Girvan Newman community detection on case study network.\label{fig:CaseStudyGN}}  
\end{figure}

In Figure \ref{fig:CaseStudyGN}$(a)$ we show the case study network after the GN algorithm has been executed and 16 communities have been detected. 
The assignment of each node to a community is visually encoded by the different colors used to depict the nodes.
To improve the clarity of the network visualization, we exploit the clustered view as shown in Figure \ref{fig:CaseStudyGN}$(b)$. This configuration adopts a modified force-directed layout in which nodes of the same community (same colors) form macro-nodes visualized with a circular layout. In such a way, inter-connectivity among communities is captured better. 
The macro-nodes can be further exposed to reveal intra-community relationships (see Figure \ref{fig:CaseStudyGN}$(c)$). 

In this case we were mainly interested not only to those nodes which occupy prominent positions.
Rather we focused on those edges whose deletion during the execution of the algorithm unveils new structural configurations which in turn can be investigated using other information available to police detectives.
This analysis will result of fundamental importance for the successful outcome of the investigation.

Thus, the analysis of a criminal network can be accomplished using \textit{LogAnalysis} as follows: \emph{(i)} data extracted from heterogeneous sources must be parsed; \emph{(ii)} a mathematical model in form of a network is derived; \emph{(iii)} a node-link layout for the visual representation is chosen; \emph{(iv)} communities are detected and visualized; \emph{(v)} the member of each cluster is analyzed in depth and, \emph{(vi}) step \emph{(iv)} is refined using the results of step \emph{(v)}.

The choice of the best level of granularity during the clustering is not automatic.
In this case study it was derived from Table \ref{table:cluster_step}, where are shown the edges which were deleted and the number of clusters which were obtained step by step until the best configuration was obtained.

In Table \ref{table:cluster_step} are shown the edges and the nodes through which information can flow towards all the members of the network or, at least, a large part of it.
A detailed analysis demonstrated, however, that the more central edges are not always responsible of driving the majority of the information.
They are, of course, important edges from a topological point of view and ``lethal'' when regarded as members of a criminal network, but only on a theoretical viewpoint.
An important consideration follows: the algorithms of clustering, when used to analyze a criminal network help to detect the most close groups of the network, but the nature of the relations must be carefully evaluated using information which can not be directly drawn from the mathematical model or its graphical representation.
The Social Network Analysis applied to our case study for example shows that the node with the highest degree (i.e., the highest number of phone calls) has a lower betweenness centrality if compared to other nodes.
In fact, criminal networks heavily employ secrecy to escape investigations and, in particular, a policy of internal communications according to which the most important members issue orders to a very limited number of members which in turn make them known to an increasing number of less important members until the leaves of the network are informed.

\begin{center}
\begin{table}[!h]
	    \caption{Results of the application of the GN algorithm to the case study. Are shown the edges which were deleted at each iteration of the EdgeBetweennessClusterer algorithm along with the incident nodes. Are also shown the edges through which information can still flow towards all the members of the network or, at least, a large part of it.}
{\footnotesize
\begin{tabular}{clcc}
\toprule
\multicolumn{1}{c}{N. Clusters} & N. & Edges & Vertices \\
\midrule
\multirow{8}*{2 Clusters} & 1 & 634 & 25$\longleftrightarrow$1 \\
\cmidrule(l){2-4}
& 2 & 576 & 64$\longleftrightarrow$44 \\
\cmidrule(l){2-4}
& 3 & 635 & 25$\longleftrightarrow$33 \\
\cmidrule(l){2-4}
& 4 & 679 & 5$\longleftrightarrow$1 \\
\cmidrule(l){2-4}
& 5 & 651 & 1$\longleftrightarrow$19 \\
\cmidrule(l){2-4}
& 6 & 617 & 25$\longleftrightarrow$42 \\
\cmidrule(l){2-4}
& 7 & 614 & 43$\longleftrightarrow$23 \\
\cmidrule(l){2-4}
& 8 & 615 & 25$\longleftrightarrow$43 \\
\midrule
\multirow{7}*{3 Clusters} & 9 & 254 & 220$\longleftrightarrow$227 \\
\cmidrule(l){2-4}
& 10 & 301 & 220$\longleftrightarrow$169 \\
\cmidrule(l){2-4}
& 11 & 381 & 169$\longleftrightarrow$64 \\
\cmidrule(l){2-4}
& 12 & 681 & 1$\longleftrightarrow$4 \\
\cmidrule(l){2-4}
& 13 & 567 & 71$\longleftrightarrow$4 \\
\cmidrule(l){2-4}
& 14 & 559 & 1$\longleftrightarrow$77 \\
\cmidrule(l){2-4}
& 15 & 616 & 1$\longleftrightarrow$42 \\
\midrule
\multirow{2}*{4 Clusters} & 16 & 610 & 20$\longleftrightarrow$44 \\
\cmidrule(l){2-4}
& 17 & 612 & 44$\longleftrightarrow$20 \\
\midrule
\multirow{6}*{5 Clusters} & 18 & 638 & 1$\longleftrightarrow$23 \\
\cmidrule(l){2-4}
& 19 & 17 & 368$\longleftrightarrow$1 \\
\cmidrule(l){2-4}
& 20 & 306 & 1$\longleftrightarrow$219 \\
\cmidrule(l){2-4}
& 21 & 104 & 304$\longleftrightarrow$1 \\
\cmidrule(l){2-4}
& 22 & 300 & 220$\longleftrightarrow$4 \\
\cmidrule(l){2-4}
& 23 & 299 & 220$\longleftrightarrow$1 \\
\bottomrule
\end{tabular}
}
\quad \quad
{\footnotesize
\begin{tabular}{clcc}
\toprule
\multicolumn{1}{c}{N. Clusters} & N. & Edges & Vertices \\
\midrule
\multirow{6}*{6 Clusters} & 24 & 639 & 23$\longleftrightarrow$4 \\
\cmidrule(l){2-4}
& 25 & 687 & 81$\longleftrightarrow$4 \\
\cmidrule(l){2-4}
& 26 & 16 & 368$\longleftrightarrow$4 \\
\cmidrule(l){2-4}
& 27 & 304 & 219$\longleftrightarrow$23 \\
\cmidrule(l){2-4}
& 28 & 641 & 22$\longleftrightarrow$23 \\
\midrule
\multirow{2}*{7 Clusters} & 29 & 611 & 44$\longleftrightarrow$25 \\
\cmidrule(l){2-4}
& 30 & 601 & 50$\longleftrightarrow$44 \\
\midrule
\multirow{6}*{8 Clusters} & 31 & 275 & 222$\longleftrightarrow$227 \\
\cmidrule(l){2-4}
& 32 & 255 & 227$\longleftrightarrow$226 \\
\cmidrule(l){2-4}
& 33 & 274 & 223$\longleftrightarrow$227 \\
\cmidrule(l){2-4}
& 34 & 276 & 221$\longleftrightarrow$227 \\
\cmidrule(l){2-4}
& 35 & 273 & 224$\longleftrightarrow$227 \\
\cmidrule(l){2-4}
& 36 & 272 & 225$\longleftrightarrow$227 \\
\midrule
9 Clusters & 37 & 569 & 64$\longleftrightarrow$71 \\
\midrule
\multirow{10}*{10 Clusters} & 38 & 341 & 169$\longleftrightarrow$199 \\
\cmidrule(l){2-4}
& 39 & 350 & 169$\longleftrightarrow$193 \\
\cmidrule(l){2-4}
& 40 & 347 & 169$\longleftrightarrow$195 \\
\cmidrule(l){2-4}
& 41 & 344 & 169$\longleftrightarrow$197 \\
\cmidrule(l){2-4}
& 42 & 372 & 169$\longleftrightarrow$177 \\
\cmidrule(l){2-4}
& 43 & 375 & 169$\longleftrightarrow$175 \\
\cmidrule(l){2-4}
& 44 & 369 & 169$\longleftrightarrow$179 \\
\cmidrule(l){2-4}
& 45 & 356 & 169$\longleftrightarrow$189 \\
\cmidrule(l){2-4}
& 46 & 359 & 169$\longleftrightarrow$187 \\
\bottomrule
\end{tabular}
}
\label{table:cluster_step}
\end{table}
\end{center}

In our case study, the nodes having the highest number of communications (i.e., the highest degree) represent the lieutenants of the criminal organization and not necessarily the boss of the clan, while the edges traversed by the highest number of shortest paths (i.e., having the highest betweenness centrality) represent the most important links among the various groups.

Moreover, the granularity of the clustering allows to identify the members and the edges which represent the ideal target when trying to hinder the criminal activities of the clan.

\begin{figure}[!h]
 \begin{minipage}[t]{.665\linewidth}\vspace{0pt}
 	\subfloat[][\emph{Case study network after applying the Newman algorithm. 19 communities have been detected, 5 of which are visualized, centered on the selected node ("Tobias", in red)}.]
	{\includegraphics[width=0.97\columnwidth]{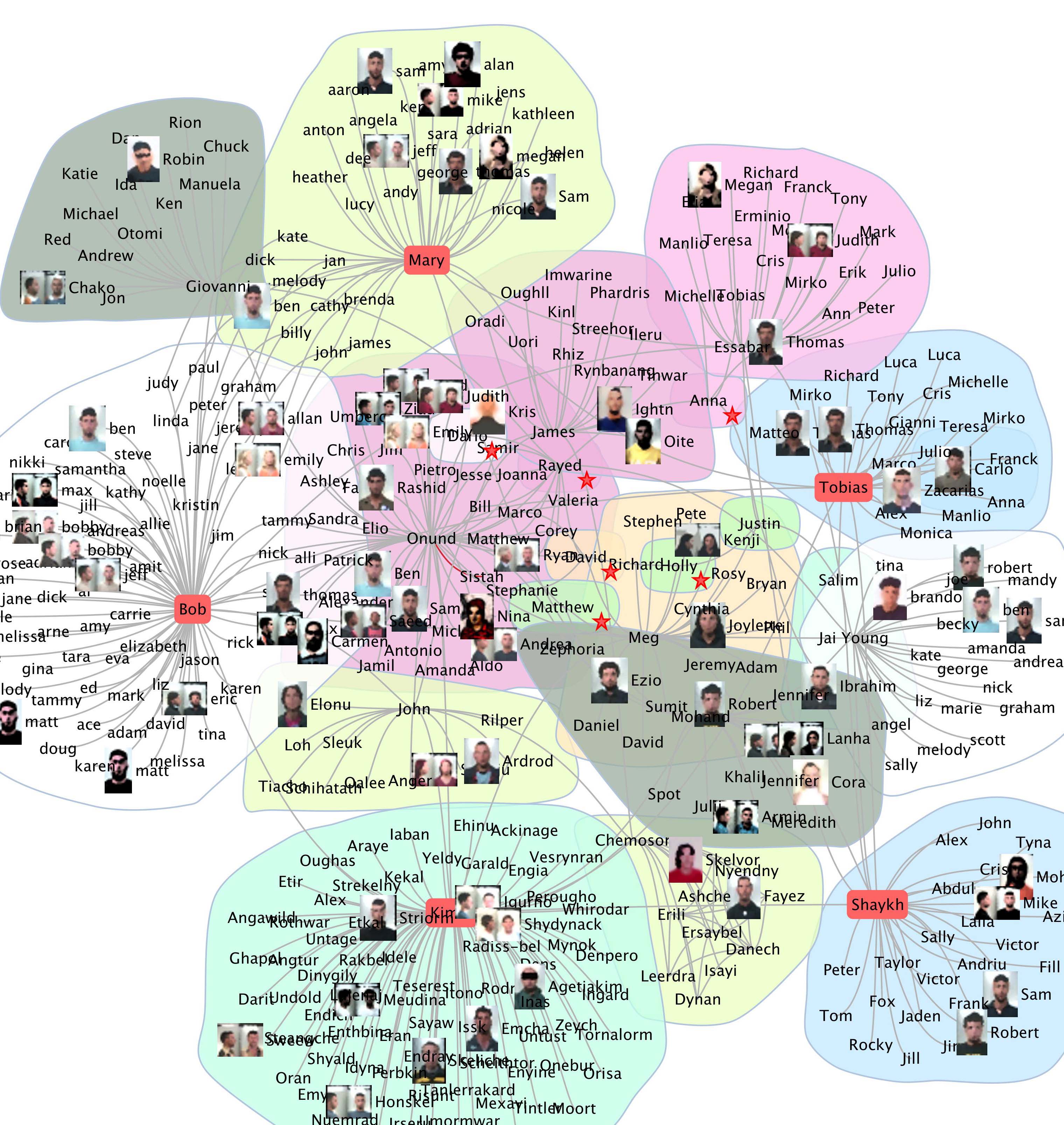}}
 \end{minipage}
 \hfill
 \begin{minipage}[t]{.335\linewidth}\vspace{0pt}\raggedright
  \begin{minipage}[t]{\linewidth}\vspace{0pt}\raggedright
	\subfloat[][\emph{Filtered communities with intercommunity hops 2 from Tobias node}.]
	{\includegraphics[width=.99\columnwidth]{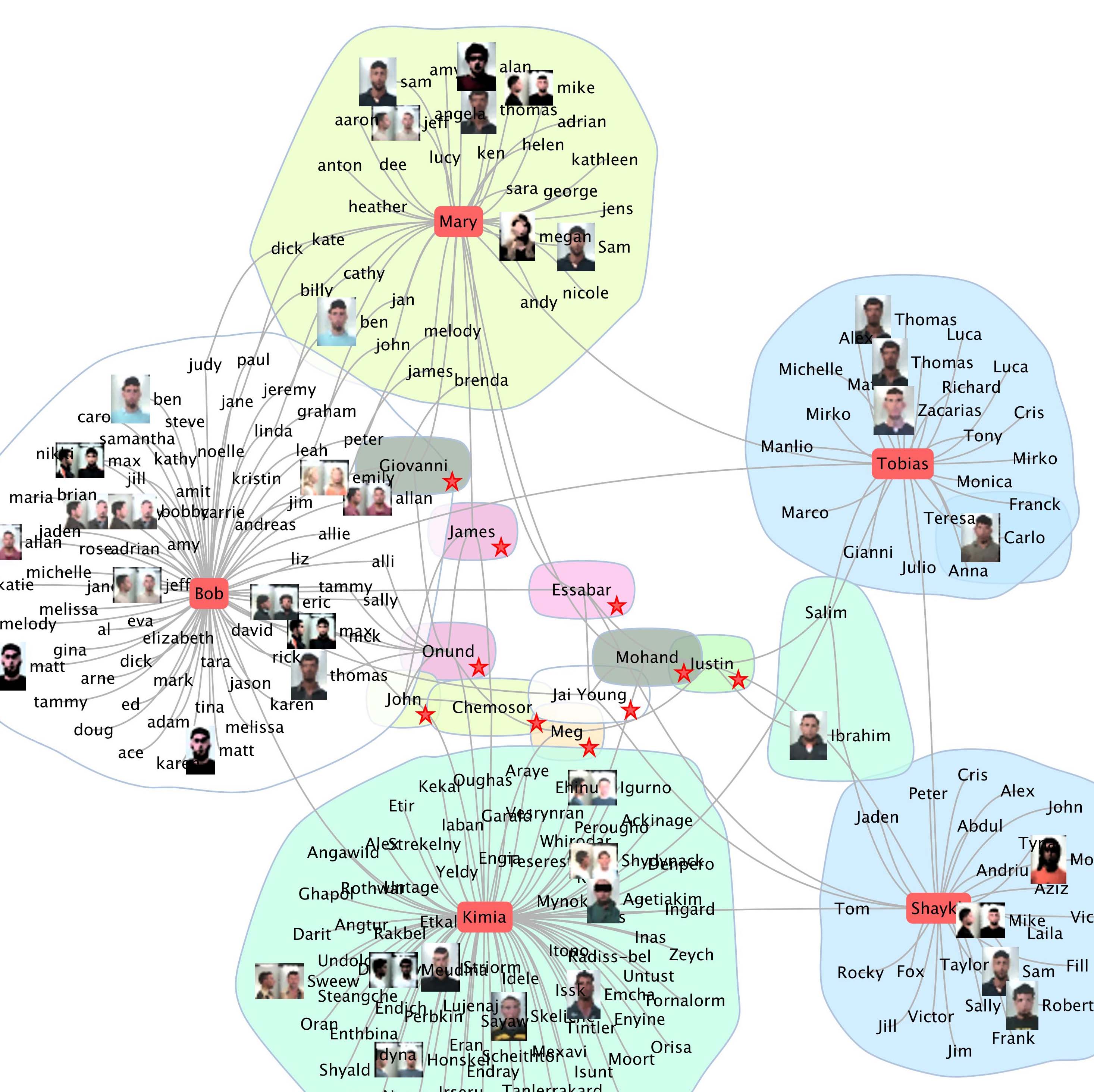}} \\
	\subfloat[][\emph{Egonet del nodo Tobias (intecommunity hops = 1)}.]
	{\includegraphics[width=1.0\columnwidth]{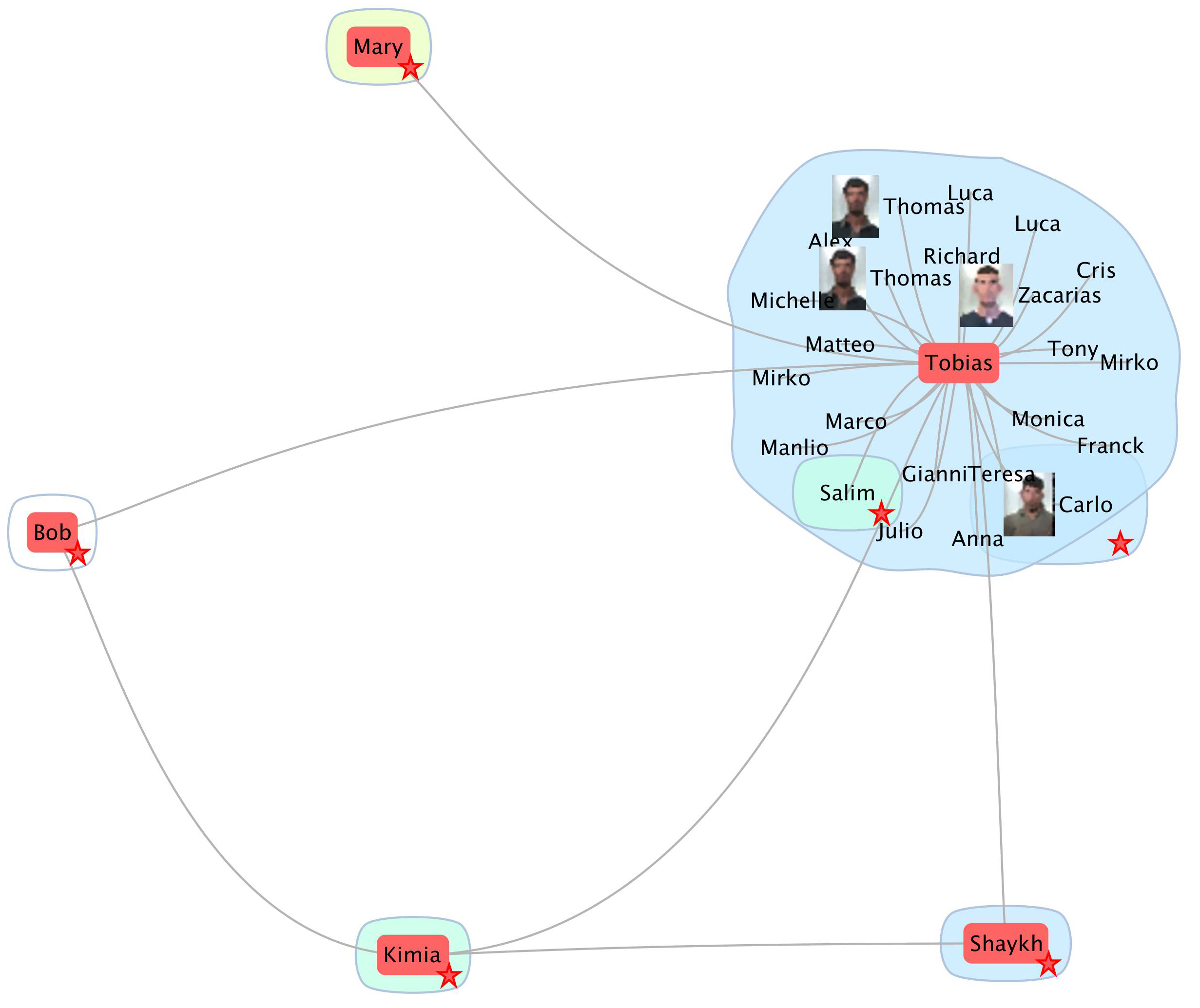}}
  \end{minipage}
  \begin{minipage}[t]{\linewidth}\vspace{20pt}\raggedright
	\end{minipage}
\end{minipage}
 \caption{Girvan Newman community detection on case study network.\label{fig:CaseStudyNewman}}  
\end{figure}

The next step of analysis is carried out by using the Newman algorithm. Figure \ref{fig:CaseStudyNewman}(a) shows communities embedded in convex hulls. Since the visualization might be cluttered and compromise the interpretation of the results, we here exploited the community compression techniques described above to improve the quality of the representation. For example, by setting the inter-community hop filter to a value of 2, Figure \ref{fig:CaseStudyNewman}(b) shows the communities, and the respective members, that can be reached from the selected nodes at most in two hops. 
Figure \ref{fig:CaseStudyNewman}(c) represents the egonet of the selected user, and the summary of communities connected in one hop. 

\subsection{Overlapping communities}
As already discussed in Section \ref{sec:related}, an important aspect in the analysis of communities is represented by the potential overlap of communities.
Both the algorithms implemented in \textit{LogAnalysis} actually perform a partition of the network, thus assigning each of the the nodes to exactly one cluster.
Often this is not a correct representation, at least on a semantic basis, of the network.
In a specific case such ours, even the algorithmic approaches described in \cite{palla2005uncovering,sun2011identification} may produce questionable results because of the multiplicity of meanings which can be given to any edge of the network.
For this reasons, we decided to implement \textit{LogAnalysis} in such a way which allows the user to choice the level of clustering in order to approximate the results.
The network shown in Figure \ref{fig:EsempioClustering} is an example of the level of clustering we believed to be the most appropriate according to the aforesaid criteria.

\begin{figure}[!h]
\centering
\subfloat[][\emph{28 edges deleted}.]
{\includegraphics[width=.44\columnwidth]{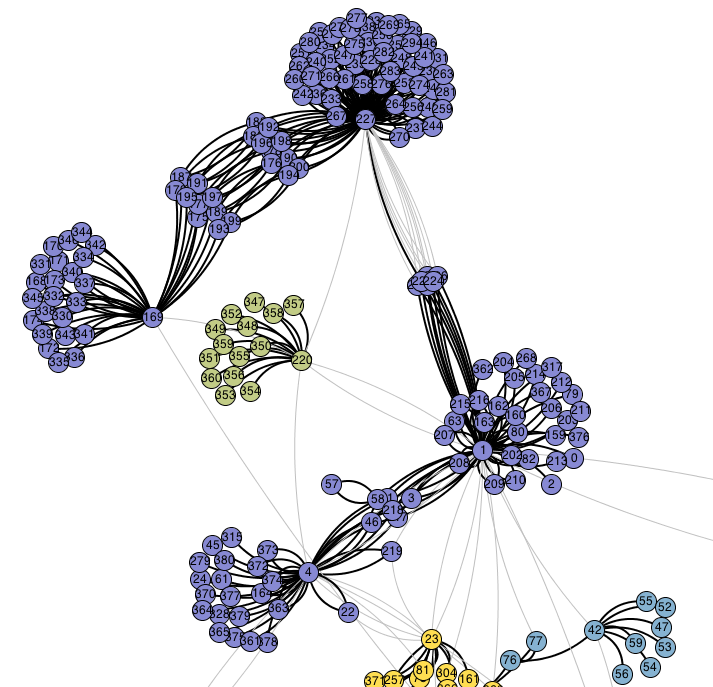}} 
\subfloat[][\emph{30 edges deleted}.]
{\includegraphics[width=.44\columnwidth]{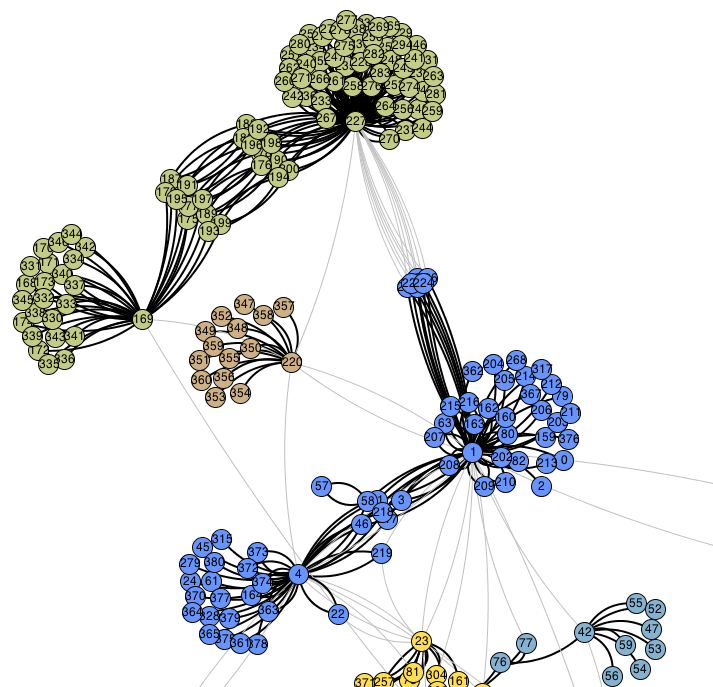}} \\
\subfloat[][\emph{36 edges deleted}.]
{\includegraphics[width=.44\columnwidth]{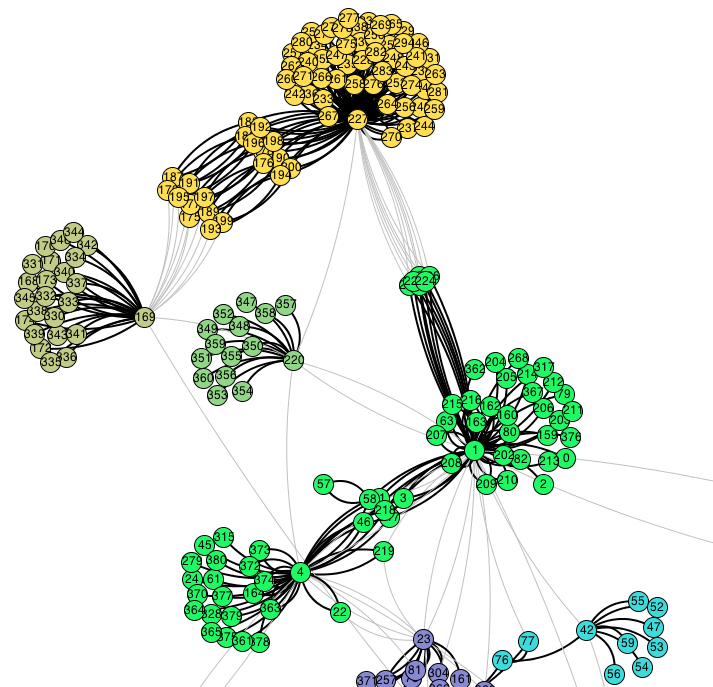}} 
\subfloat[][\emph{47 edges deleted}.]
{\includegraphics[width=.44\columnwidth]{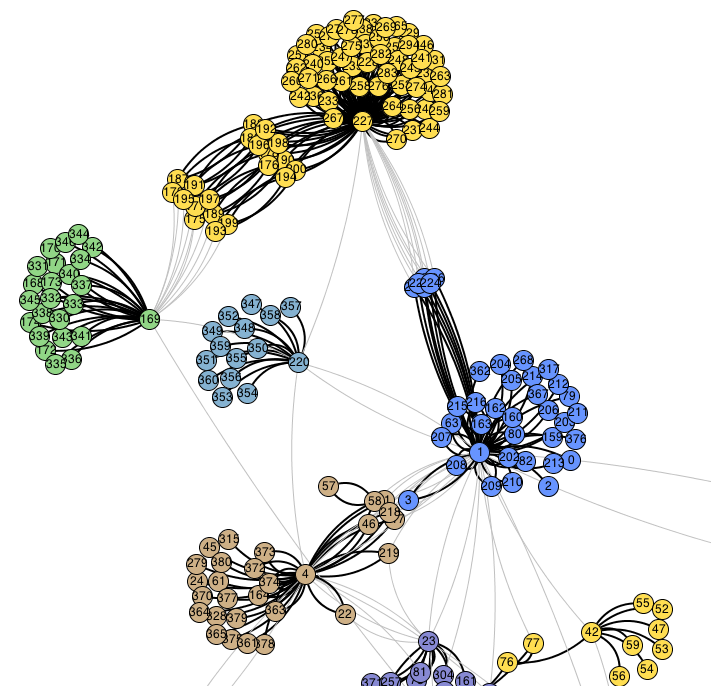}}
\caption{Non coherent examples of clustering produced applying the GN clustering algorithm \cite{girvan2002community}.}
\label{fig:OverlapGN}
\end{figure}

Some examples follow.
Figure \ref{fig:OverlapGN} shows a situation in which the GN algorithm can produce a series of results in which the outcoming partition does not correspond to the results of the studied network.
In the example, only a portion of the entire network is shown.
After the deletion of 28 edges (see Figure \ref{fig:OverlapGN}(a)), a community is obtained (colored in violet) which is composed of more groups.
The deletion of two more edges leads to the configuration shown in Figure \ref{fig:OverlapGN}(b).
Some of the nodes belonging to the blue cluster should belong also to the green cluster.
In this example, even if the clustering obtained through the application of the edge betweenness centrality is undoubtedly correct from the computational point of view, nonetheless is debatable from the semantic point of view.
Our conclusion is that in such situations an automatic computation should be supported by the assistance of the analyst.
Other examples are shown in Figures \ref{fig:OverlapGN}(c) and \ref{fig:OverlapGN}(d).
In these cases we have the same results when partitioning the nodes among yellow and green cluster (see Figure \ref{fig:OverlapGN}(c)) and when partitioning the nodes among blue and brown cluster (see Figure \ref{fig:OverlapGN}(d)).
In each of the aforesaid examples, the interconnected nodes could belong to one or the other group or both, or more simply they could belong to a group of its own which has very few links to other groups.

\begin{figure}[!h]
\centering
\subfloat[][\emph{$Q_{max}$ modularity}.]
{\includegraphics[width=1.0\columnwidth]{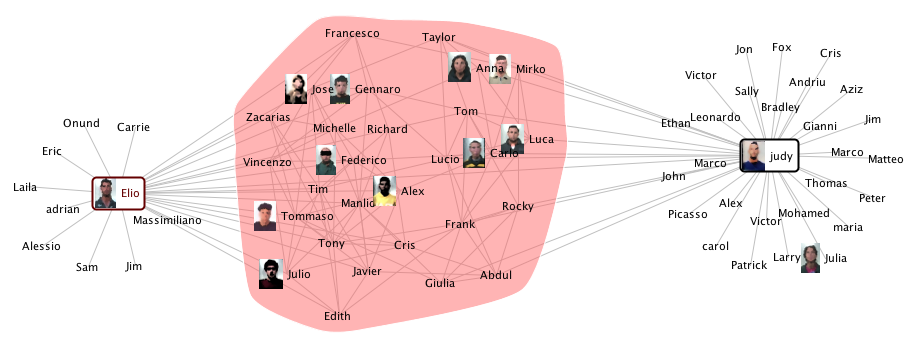}} \\
\subfloat[][\emph{$Q_{max} - 1$ modularity}.]
{\includegraphics[width=1.0\columnwidth]{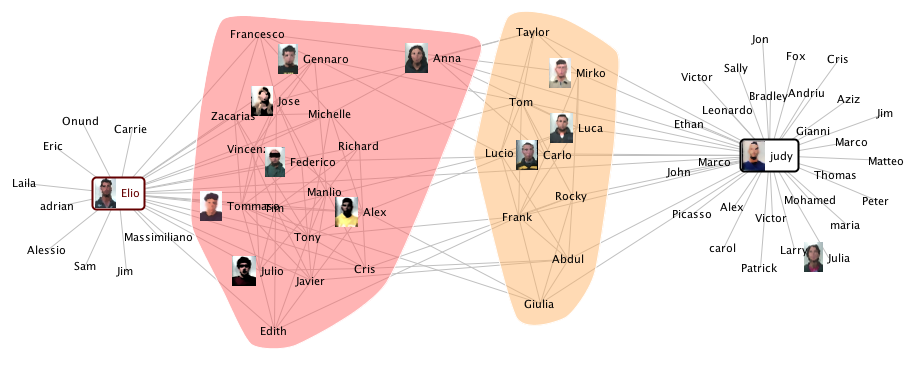}}
\caption{An example of community detection using the Newman algortihm \cite{newman2004fast}. The convex-hull layout has been adopted for the visualization of the communities.} 
\label{fig:NewmanFast1}
\end{figure}

\begin{figure}[!h]
\centering
\subfloat[][\emph{The criminal network at time $t_1$}.]
{\includegraphics[width=1.0\columnwidth]{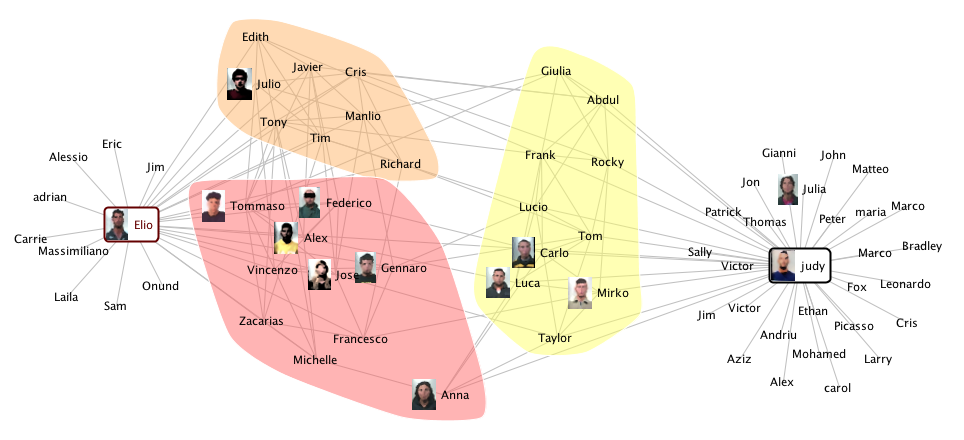}} \\
\subfloat[][\emph{The criminal network at time $t_2$}.]
{\includegraphics[width=1.0\columnwidth]{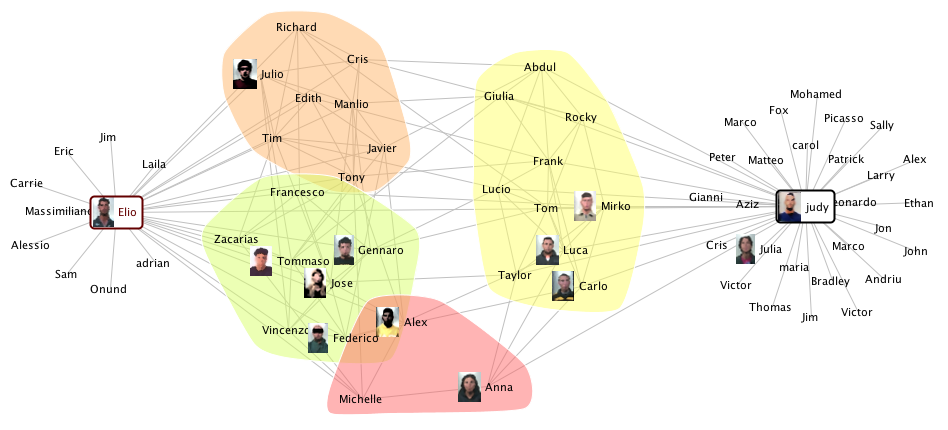}}
\caption{Community detection of a time-varying criminal network.}
\label{fig:NewmanFast2}
\end{figure}

Also the interconnections among various communities have been analyzed using the Newman community detection algorithm \cite{newman2004fast}. 
Figure \ref{fig:NewmanFast1} shows the initial phase of the execution of the algorithm.
In Figure \ref{fig:NewmanFast1}(a) only one cluster has been detected which is composed of the nodes interconnected among the external clusters represented by the nodes ``Elio'' and ``Judy'', while in Figure \ref{fig:NewmanFast1}(b), $Q_{max}$ has been interactively decreased to a previous lower value.
As a consequence, the interconnected nodes are subdivided and new communities emerge.

The in-depth analysis carried out on the members of the clusters interconnected shown in Figure \ref{fig:NewmanFast1} and the temporal analysis accomplished with \textit{LogAnalysis} allowed the investigators to discover that some clans belonging to the criminal network had worked with a certain degree of autonomy and were responsible of some murders.
It turned out from the investigations that these clans had the task of committing murders. 
In Figure \ref{fig:NewmanFast2} are shown the clans at times $t_1$ and $t_2$.

\begin{figure}[!h]
\centering
\subfloat[][\emph{Dendrogram: 428 edges,  381 nodes, 14 clusters, 84 nodes in the largest cluster}.]
{\includegraphics[width=1.00\columnwidth]{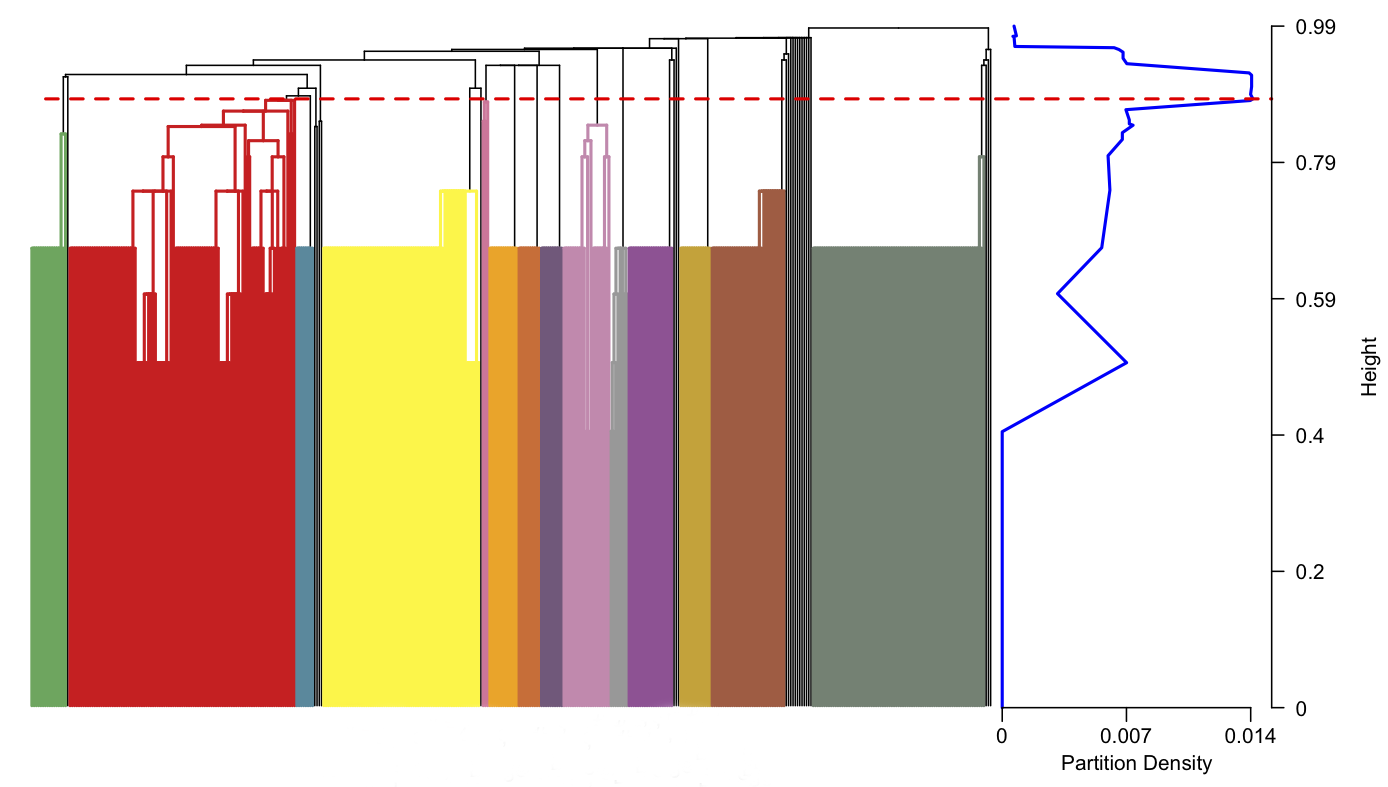}} \\ 
\subfloat[][\emph{Community membership matrix for the most connected nodes. Colors indicate community-specific membership}.]
{\includegraphics[width=0.46\columnwidth]{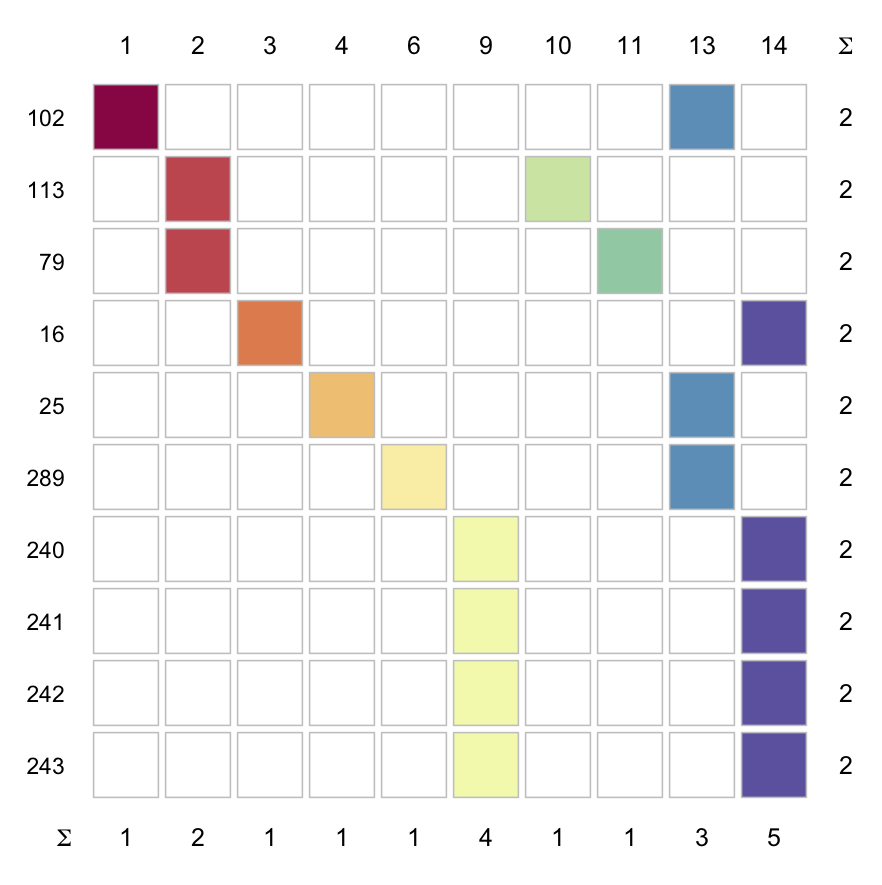}} \quad
\subfloat[][\emph{Modularity of various communities}.]
{\includegraphics[width=0.50\columnwidth]{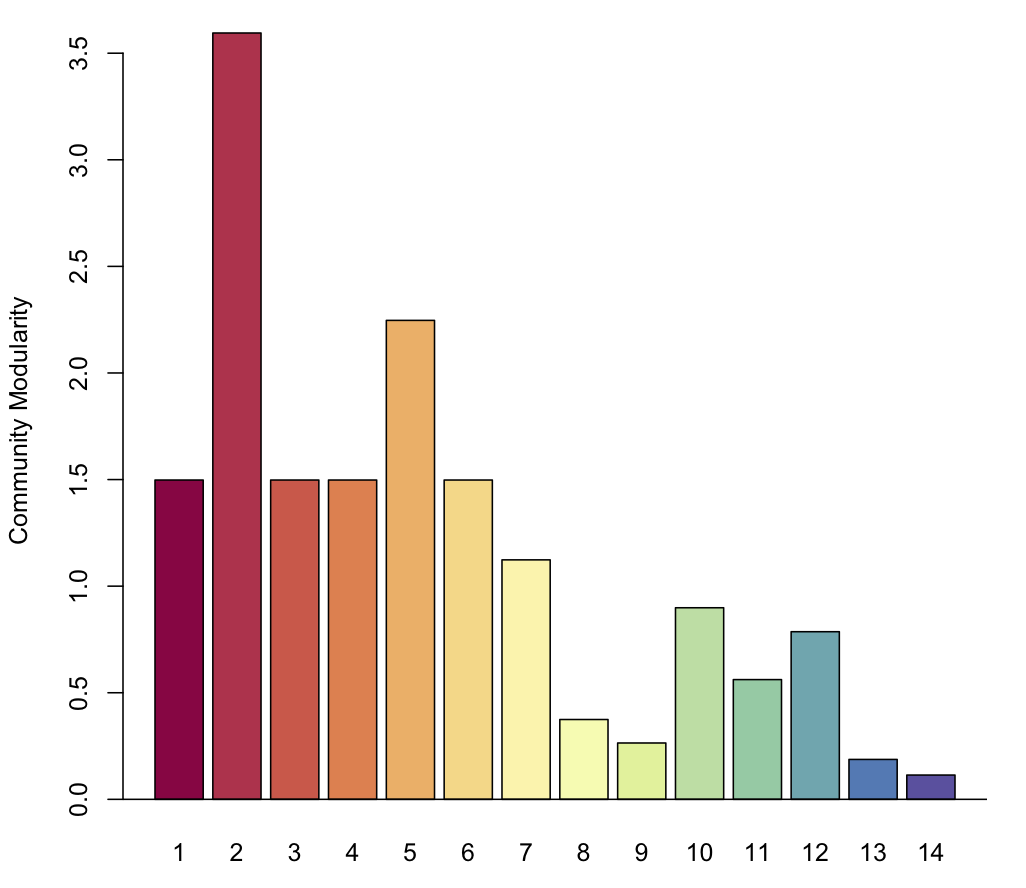}}
\caption{The figure shows: (a) the dendrogram resulting from the community detection; (b) the community membership matrix for the most connected nodes; (c) the distribution of modularity for the clustering resulting from the dendrogram cut of subfigure (a). Colors in the membership matrix correspond to those of the histogram in subfigure (c).}
\label{fig:dendrogram}
\end{figure}

Some additional remarks follow.
Applying Newman community detection algorithm with an automatic clustering produces a partition  according to which the criminal network is composed of $14$ clusters.
This can be seen from the dendrogram shown in Figure \ref{fig:dendrogram}.
The maximum partition density is $0.014$ and the largest community is composed of $84$ nodes.
As already seen, this clustering is not coherent with the real structural subdivision of the criminal network as it emerged from the supervised interactive community detection combined with additional comparisons and in-depth examinations obtained from other informative sources.
Nevertheless this result was very interesting in that important information regarding some members of the network emerged.

In particular, from the analysis of the different levels of clustering interactively selected, and from the observation of the relative variations in the obtained configurations, we identified which elements of the network were affected mostly.

This is shown in Figure \ref{fig:dendrogram}(b), with respect to the most connected nodes and how they belong to the $14$ different communities.
For example, node $102$ belongs to clusters $1$ and $13$, while the most important nodes belonging to cluster $14$ are 16, 240, 241, 242 and 243.

We also computed the modularity of the various communities, that is a measure of how dense are the connections among the nodes within the clusters in respect to the connections between nodes in different clusters.
Figure \ref{fig:dendrogram}(c) shows the modularity of each cluster of the criminal network.

\begin{figure}[!h]
\centering
{\includegraphics[width=0.85\columnwidth]{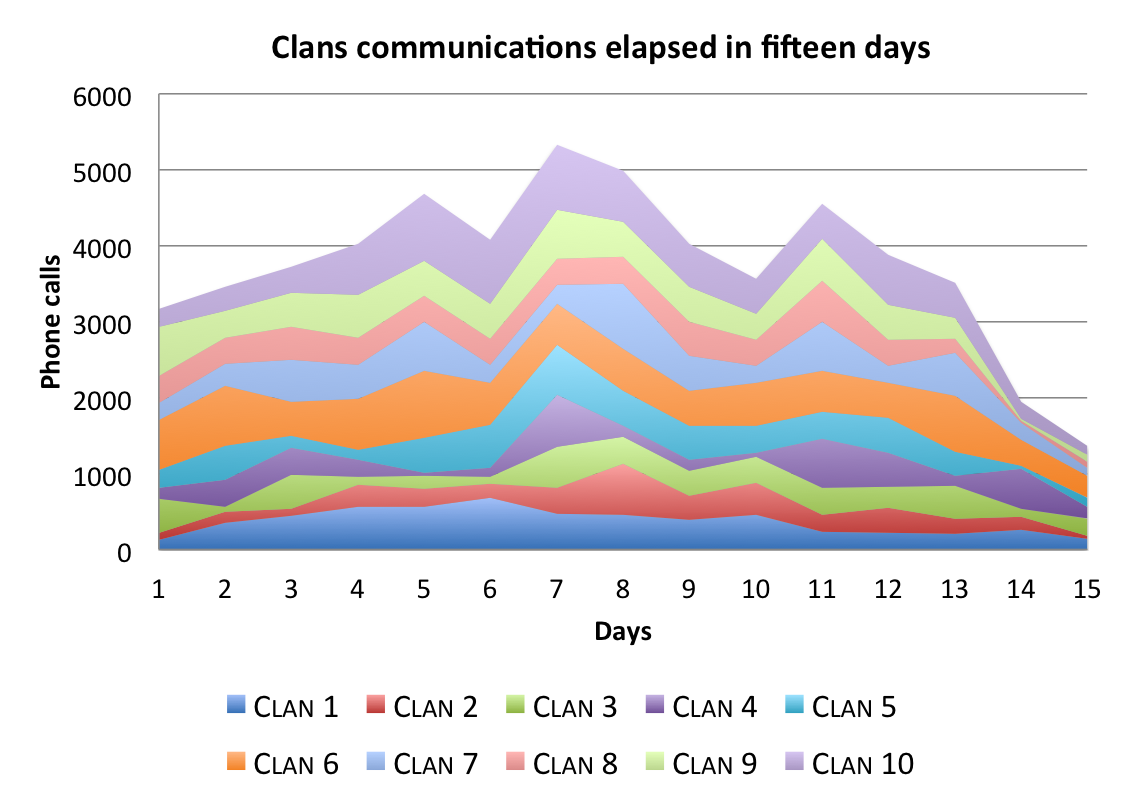}}
\caption{Stacked histogram showing the phone call traffic carried out by each community in the time interval of 15 days.}
\label{fig:stacked}
\end{figure}

The analysis of the distribution of phone calls carried out by each ``clan'' is a method generally very useful it must be decided if a good level of clustering has been obtained after the execution of the community  detection algorithm (see Figure \ref{fig:stacked}). 
The goal of this analysis is twofold: first, it identifies the groups among which the largest number of phone calls, texts, MMS took place, second, it highlights the peaks of the stream of communications related not to single users but rather to each cluster as a whole, on the occasion of a crime.
 
\section{Conclusions} \label{sec:conclusions}

In the latest decade or so there has been an active involvement of academic researchers in the study of terrorist and criminal networks to improve public safety. 
In this respect, Social Network Analysis has proved a valuable tool in order to ascertain the central members of criminal networks, the existence of subgroups, the interactions among individuals and subgroups, the flow of information in the network, the sensitive members and/or relations whose removal could eventually lead to the destruction of the network.
  
In this context, the analysis of phone call networks is crucial to gain fundamental information about inter-connectivity and communication among criminals, and to progress fruitfully the investigations.
The study of information flow allows to identify those individuals who play a key role inside the criminal organization, or connect different subgroups.
Statistical approaches also provide remarkable insights, for example if one considers the quantity of information and their temporal distribution with reference to a given criminal act.
Moreover, the spatial distribution of such events can be taken into account, since it gives insights with respect to the identification of suspected individuals and their most frequent locations.

In this work we presented \textit{LogAnalysis}, an expert system that allows for semi-supervised detection of criminal communities in networks reconstructed from phone call records. We discussed some of its features describing how they are instrumental to study criminal networks, presenting a case study inspired by a real criminal investigation. This allowed us to unveil few  primary characteristics of criminal communities in real world phone call networks.

The analysis of criminal networks cannot be reduced, however, to the study of the relations established by means of phone calls.
We must take into due account a larger amount of data, possibly originated from various sources. 
This is the case, for example, of physical meetings and financial transactions.
Also time plays an important role, in that relations and transactions usually may or may not happen simultaneously.

To analyze such types of data a radical extension of the capabilities of \textit{LogAnalysis} is necessary.
For this reason we are designing a natural successor of \textit{LogAnalysis}, conceived to study multiplex and temporal criminal networks. Such expert system will integrate and deal with multiple data sources including online social network data and financial records, and it will be integrated with other law enforcement databases to infer and learn new associations in a fully unsupervised way.

\section*{References}
\bibliography{sigproc}

\begin{thebibliography}{10}
\expandafter\ifx\csname url\endcsname\relax
  \def\url#1{\texttt{#1}}\fi
\expandafter\ifx\csname urlprefix\endcsname\relax\def\urlprefix{URL }\fi
\expandafter\ifx\csname href\endcsname\relax
  \def\href#1#2{#2} \def\path#1{#1}\fi

\bibitem{onnela2007analysis}
J.-P. Onnela, J.~Saram{\"a}ki, J.~Hyv{\"o}nen, G.~Szab{\'o}, M.~A. De~Menezes,
  K.~Kaski, A.-L. Barab{\'a}si, J.~Kert{\'e}sz, Analysis of a large-scale
  weighted network of one-to-one human communication, New Journal of Physics
  9~(6) (2007) 179.

\bibitem{onnela2007structure}
J.~Onnela, J.~Saram{\"a}ki, J.~Hyv{\"o}nen, G.~Szab{\'o}, D.~Lazer, K.~Kaski,
  J.~Kert{\'e}sz, A.~Barab{\'a}si, Structure and tie strengths in mobile
  communication networks, Proc. Natl. Acad. Sci. 104~(18) (2007) 7332.

\bibitem{candia2008uncovering}
J.~Candia, M.~Gonz{\'a}lez, P.~Wang, T.~Schoenharl, G.~Madey, A.~Barab{\'a}si,
  Uncovering individual and collective human dynamics from mobile phone
  records, Journal of Physics A: Mathematical and Theoretical 41 (2008) 224015.

\bibitem{eagle2008mobile}
N.~Eagle, A.~Pentland, D.~Lazer, Mobile phone data for inferring social network
  structure, Social Computing, Behavioral Modeling, and Prediction (2008)
  79--88.

\bibitem{eagle2009inferring}
N.~Eagle, A.~Pentland, D.~Lazer, Inferring friendship network structure by
  using mobile phone data, Proc. Natl. Acad. Sci. 106~(36) (2009) 15274.

\bibitem{becker2013human}
R.~Becker, R.~C{\'a}ceres, K.~Hanson, S.~Isaacman, J.~M. Loh, M.~Martonosi,
  J.~Rowland, S.~Urbanek, A.~Varshavsky, C.~Volinsky, Human mobility
  characterization from cellular network data, Communications of the ACM 56~(1)
  (2013) 74--82.

\bibitem{Ahn2007}
Y.~Ahn, S.~Han, H.~Kwak, S.~Moon, H.~Jeong, {Analysis of topological
  characteristics of huge online social networking services}, in: Proceedings
  of the 16th international conference on World Wide Web, ACM, 2007, pp.
  835--844.

\bibitem{Benevenuto2009}
F.~Benevenuto, T.~Rodrigues, M.~Cha, V.~Almeida, {Characterizing user behavior
  in online social networks}, in: Proceedings of the 9th ACM SIGCOMM conference
  on Internet measurement conference, ACM, 2009, pp. 49--62.

\bibitem{catanese2011crawling}
S.~A. Catanese, P.~De~Meo, E.~Ferrara, G.~Fiumara, A.~Provetti, Crawling
  {F}acebook for social network analysis purposes, in: Proceedings of the
  international conference on web intelligence, mining and semantics, ACM,
  2011, p.~52.

\bibitem{catanese2012extraction}
S.~Catanese, P.~De~Meo, E.~Ferrara, G.~Fiumara, A.~Provetti, Extraction and
  analysis of {F}acebook friendship relations, in: Computational Social
  Networks, Springer, 2012, pp. 291--324.

\bibitem{conover2013geospatial}
M.~D. Conover, C.~Davis, E.~Ferrara, K.~McKelvey, F.~Menczer, A.~Flammini, The
  geospatial characteristics of a social movement communication network, PloS
  one 8~(3) (2013) e55957.

\bibitem{conover2013digital}
M.~D. Conover, E.~Ferrara, F.~Menczer, A.~Flammini, The digital evolution of
  occupy wall street, PloS one 8~(5) (2013) e64679.

\bibitem{xu2005criminal}
J.~Xu, H.~Chen, Criminal network analysis and visualization, Communications of
  the ACM 48~(6) (2005) 100--107.

\bibitem{morselli2010assessing}
C.~Morselli, Assessing vulnerable and strategic positions in a criminal
  network, Journal of Contemporary Criminal Justice 26~(4) (2010) 382--392.

\bibitem{chen2003coplink}
H.~Chen, D.~Zeng, H.~Atabakhsh, W.~Wyzga, J.~Schroeder, Coplink: managing law
  enforcement data and knowledge, Communications of the ACM 46~(1) (2003)
  28--34.

\bibitem{xu2005crimenet}
J.~J. Xu, H.~Chen, Crimenet explorer: a framework for criminal network
  knowledge discovery, ACM Transactions on Information Systems 23~(2) (2005)
  201--226.

\bibitem{Brian2006}
W.~Wright, D.~Schroh, P.~Proulx, A.~Skaburskis, B.~Cort,
  \href{http://doi.acm.org/10.1145/1124772.1124890}{The sandbox for analysis:
  Concepts and methods}, in: Proceedings of the SIGCHI Conference on Human
  Factors in Computing Systems, CHI '06, ACM, New York, NY, USA, 2006, pp.
  801--810.
\newblock \href {http://dx.doi.org/10.1145/1124772.1124890}
  {\path{doi:10.1145/1124772.1124890}}.
\newline\urlprefix\url{http://doi.acm.org/10.1145/1124772.1124890}

\bibitem{Everett2006}
N.~J. Pioch, J.~O. Everett, Polestar: collaborative knowledge management and
  sensemaking tools for intelligence analysts., in: P.~S. Yu, V.~J. Tsotras,
  E.~A. Fox, B.~L. 0001 (Eds.), CIKM, ACM, 2006, pp. 513--521.

\bibitem{McAndrew99}
D.~McAndrew, The structural analysis of criminal networks, D. Canter and L.
  Alison Eds. The Social Psychology of Crime: Groups, Teams, and Networks,
  Offender Profiling Series 3.

\bibitem{milgram1967small}
S.~Milgram, {The small world problem}, Psychology today 2~(1) (1967) 60--67.

\bibitem{Travers1969}
J.~Travers, S.~Milgram, {An experimental study of the small world problem},
  Sociometry 32~(4) (1969) 425--443.

\bibitem{Zachary1980}
W.~Zachary, {An information flow model for conflict and fission in small
  groups}, Journal of Anthropological Research 33~(4) (1977) 452--473.

\bibitem{albert1999diameter}
R.~Albert, H.~Jeong, A.~Barabasi, {The diameter of the world wide web}, Nature
  401 (1999) 130--131.

\bibitem{albert2002statistical}
R.~Albert, A.~Barab{\'a}si, {Statistical mechanics of complex networks},
  Reviews of modern physics 74~(1) (2002) 47--97.

\bibitem{ferrara2011topological}
E.~Ferrara, G.~Fiumara, Topological features of online social networks,
  Communications on Applied and Industrial Mathematics 2~(2) (2011) 15--33.

\bibitem{sparrow91}
M.~K. Sparrow, The application of network analysis to criminal intelligence: An
  assessment of the prospects, Social Networks 13~(3) (1991) 251--274.

\bibitem{BakerFaulkner93}
W.~Baker, R.~Faulkner, The social organization of conspiracy: illegal networks
  in the heavy electrical equipment industry, Am. Social. Rev. 58 (1993)
  837--860.

\bibitem{Klerks01thenetwork}
P.~Klerks, E.~Smeets, The network paradigm applied to criminal organizations:
  Theoretical nitpicking or a relevant doctrine for investigators? recent
  developments in the netherlands, Connections 24 (2001) 53--65.

\bibitem{Renfro01}
R.~S. Renfro, R.~F. Deckro, A social network analysis of the iranian
  government, 69th Military Operational Research Symposium.

\bibitem{Silke01}
A.~Slike, The devil you know: Continuing problems with research on terrorism,
  Terrorism and Political Violence 13 (2001) 1--14.

\bibitem{Brannan01}
D.~W. Brannan, P.~F. Esler, N.~T. Anders~Strindberg, Talking to terrorists:
  Towards an independent analytical framework for the study of violent substate
  activism, Studies in Conflict and Terrorism 24~(1) (2001) 3--24.

\bibitem{Arquilla01}
J.~Arquilla, D.~Ronfeldt, Networks and netwars: The future of terror, crime,
  and militancy, Survival 44~(2) (2001) 175--176.

\bibitem{Carley2006}
K.~M. Carley, Destabilization of covert networks, Comput. Math. Organ. Theory
  12~(1) (2006) 51--66.

\bibitem{Krebs02mappingnetworks}
V.~Krebs, Mapping networks of terrorist cells, Connections 24~(3) (2002)
  43--52.

\bibitem{fortunato2010community}
S.~Fortunato, Community detection in graphs, Physics Reports 486~(3-5) (2010)
  75--174.

\bibitem{girvan2002community}
M.~Girvan, M.~Newman, {Community structure in social and biological networks},
  Proc. Natl. Acad. Sci. 99~(12) (2002) 7821.

\bibitem{newman2004finding}
M.~Newman, M.~Girvan, {Finding and evaluating community structure in networks},
  Phy. Rev. E 69~(2) (2004) 26113.

\bibitem{newman2006modularity}
M.~Newman, {Modularity and community structure in networks}, PNAS 103~(23)
  (2006) 8577.

\bibitem{blondel2008fast}
V.~Blondel, J.~Guillaume, R.~Lambiotte, E.~Lefebvre, Fast unfolding of
  communities in large networks, Journal of Statistical Mechanics: Theory and
  Experiment (2008) P10008.

\bibitem{fortunato2007resolution}
S.~Fortunato, M.~Barthelemy, Resolution limit in community detection,
  Proceedings of the National Academy of Sciences 104~(1) (2007) 36--41.

\bibitem{palla2005uncovering}
G.~Palla, I.~Der{\'e}nyi, I.~Farkas, T.~Vicsek, {Uncovering the overlapping
  community structure of complex networks in nature and society}, Nature
  435~(7043) (2005) 814--818.

\bibitem{xie2011overlapping}
J.~Xie, S.~Kelley, B.~K. Szymanski, Overlapping community detection in
  networks: the state of the art and comparative study, ACM Computing Surveys
  45~(4).

\bibitem{Newman_2005}
M.~Newman, A measure of betweenness centrality based on random walks, Social
  Networks 27~(1) (2005) 39--54.

\bibitem{wasserman1994social}
S.~Wasserman, K.~Faust, Social network analysis: methods and applications,
  Cambridge University Press, 1994.

\bibitem{yee2001animated}
K.~Yee, D.~Fisher, R.~Dhamija, M.~Hearst, Animated exploration of dynamic
  graphs with radial layout, in: Proc. IEEE Symposium on Information
  Visualization, 2001, p.~43.

\bibitem{Gniadek2010}
U.~K. Wiil, J.~Gniadek, N.~Memon,
  \href{http://dblp.uni-trier.de/db/conf/asunam/asonam2010.html#WiilGM10}{Measuring
  link importance in terrorist networks.}, in: N.~Memon, R.~Alhajj (Eds.),
  ASONAM, IEEE Computer Society, 2010, pp. 225--232.
\newline\urlprefix\url{http://dblp.uni-trier.de/db/conf/asunam/asonam2010.html#WiilGM10}

\bibitem{sageman2004}
M.~Sageman, Understanding Terror Networks, University of Pennsylvania Press,
  2004.

\bibitem{Nomani2012}
M.~Todd, A.~Nomani, The Truth Left Behind: Inside the Kidnapping and Murder of
  Daniel Pearl, New York (2011) -
  http://www.publicintegrity.org/2011/01/20/2190/, 2011.

\bibitem{newman2004fast}
M.~Newman, {Fast algorithm for detecting community structure in networks},
  Phys. Rev. E 69~(6) (2004) 066133.

\bibitem{fruchterman1991graph}
T.~Fruchterman, E.~Reingold, {Graph drawing by force-directed placement},
  Software: Practice and Experience 21~(11) (1991) 1129--1164.

\bibitem{heer2005vizster}
J.~Heer, D.~Boyd, Vizster: Visualizing online social networks, in: Proc. IEEE
  Symposium on Information Visualization, 2005, p.~5.

\bibitem{sun2011identification}
P.~Sun, L.~Gao, S.~Shan~Han, Identification of overlapping and non-overlapping
  community structure by fuzzy clustering in complex networks, Information
  Sciences 181~(6) (2011) 1060--1071.

\end{thebibliography}

\end{document}